\documentclass[sigconf]{acmart}
\copyrightyear{2025}
\acmYear{2025}
\setcopyright{licensedusgovmixed}
\acmConference[SIGIR '25]{Proceedings of the 48th International ACM SIGIR Conference on Research and Development in Information Retrieval}{July 13--18, 2025}{Padua, Italy}
\acmBooktitle{Proceedings of the 48th International ACM SIGIR Conference on Research and Development in Information Retrieval (SIGIR '25), July 13--18, 2025, Padua, Italy}
\acmDOI{10.1145/3726302.3729916}
\acmISBN{979-8-4007-1592-1/2025/07}

\usepackage{graphicx} % Required for inserting images
\usepackage{algorithm}
\usepackage{algpseudocode}
\usepackage{multirow}
\usepackage{soul}
\usepackage{float}
\usepackage{afterpage}
\usepackage{caption}

\usepackage{siunitx} % Add this to your preamble

\usepackage{tcolorbox}

\DeclareMathOperator*{\argmax}{arg\,max}

\definecolor{blue}{rgb}{0.3,0.68,0.70} % a light blue color
\tcbset{
  colback=blue!5!white,    % Background color
  colframe=blue!75!black,  % Frame color
  boxsep=5pt,
  arc=4pt,
  outer arc=4pt,
  boxrule=0.5pt,
  fonttitle=\bfseries,
}
\newtheorem{theorem}{Theorem}

\title{LLM-Assisted Relevance Assessments:\\When Should We Ask LLMs for Help?}

\author{Rikiya Takehi}
\orcid{0009-0003-6336-8064}
\affiliation{
  \institution{Waseda University}
  \city{Tokyo}
  \country{Japan}
}
\email{rikiya.takehi@fuji.waseda.jp}

\author{Ellen M. Voorhees}
\orcid{0000-0002-5658-2308}
\affiliation{
  \institution{National Institute of Standards and Technology}
  \city{Gaithersburg} 
  \state{MD}
  \country{United States}
}
\email{ellen.voorhees@nist.gov}

\author{Tetsuya Sakai}
\orcid{0000-0002-6720-963X}
\affiliation{
  \institution{Waseda University}
  \city{Tokyo}
  \country{Japan}
}
\email{tetsuyasakai@acm.org}

\author{Ian Soboroff}
\orcid{0000-0003-2363-3014}
\affiliation{
  \institution{National Institute of Standards and Technology}
  \city{Gaithersburg} 
  \state{MD}
  \country{United States}
}
\email{ian.soboroff@nist.gov}

\begin{document}
\renewcommand{\shortauthors}{Rikiya Takehi, Ellen M. Voorhees, Tetsuya Sakai, \& Ian Soboroff}
\begin{abstract}
    % Test collections allow Information Retrieval (IR) researchers to quickly and easily evaluate ranking algorithms. One major problem, however, is that the process of building test collections involves significant effort in manual annotations, making its construction very costly. Consequently, test collections may become too small when under a low budget, which may lead to unstable evaluations. As a cheaper alternative, various studies have investigated the use of large language models (LLMs) to completely replace human assessors. However, while LLMs may seem to somewhat correlate with human judgments, their predictions are not perfect and often show bias. Thus a complete replacement with LLMs has recently been argued to be risky and not fully reliable.
    
    % To solve this problem, in this paper, we propose a method that effectively uses both manual annotations and LLM annotations, which helps to build a rich and reliable test collection even under a low budget. Guided by theoretical analysis, we use the LLM’s predicted relevance probabilities to select the most profitable documents to manually annotate under a budget constraint. Then, using the given manual annotations, we actively learn to calibrate the LLM predictions of the remaining documents while finding the next document to annotate manually. Experiments on TREC-7 Ad Hoc, TREC-8 Ad Hoc, TREC Robust 2004, and TREC-COVID datasets show that our method outperforms alternative solutions under almost any budget constraint.
    Test collections are information retrieval tools that allow researchers to quickly and easily evaluate ranking algorithms. While test collections have become an integral part of IR research, the process of data creation involves significant manual annotation effort, which often makes it very expensive and time-consuming. Consequently, test collections could become too small when the budget is limited, which may lead to unstable evaluations. As a cheaper alternative, recent studies have proposed the use of large language models (LLMs) to completely replace human assessors. However, while LLMs seem to somewhat correlate with human judgments, their predictions are not perfect and often show bias. Thus, a complete replacement with LLMs is argued to be too risky and not fully reliable.

    In this paper, we propose \textbf{L}LM-\textbf{A}ssisted \textbf{R}elevance \textbf{A}ssessments (\textbf{LARA})\footnote{Our code is available at \url{https://github.com/RikiyaT/LARA}}, an effective method to balance manual annotations with LLM annotations, which helps to build a rich and reliable test collection even under a low budget. We use the LLM's predicted relevance probabilities to select the most profitable documents to manually annotate under a budget constraint. With theoretical reasoning, LARA effectively guides the human annotation process by actively learning to calibrate the LLM's predicted relevance probabilities. Then, using the calibration model learned from the limited manual annotations, LARA debiases the LLM predictions to annotate the remaining non-assessed data. Experiments on TREC-7 Ad Hoc, TREC-8 Ad Hoc, TREC Robust 2004, and TREC-COVID datasets show that LARA outperforms alternative solutions under almost any budget constraint. While the community debates humans vs. LLMs in relevance assessments, we contend that, \textbf{given the same amount of human effort, it is reasonable to \textit{leverage} LLMs.}
\end{abstract}
\maketitle

\section{Introduction}
Test collections are laboratory tools that allow Information Retrieval (IR) systems to be reliably evaluated. A collection consists of relevance labels, which are annotations of whether a search result (we refer to as a \textit{document}) is relevant to the searcher's question (we refer to as a \textit{topic}). Labels may come from various sources, but the most accurate option is to use the gold assessor labels~\citep{bailey2008gold}. In test collections, these labels originate from relevance assessors who develop their own topic and annotate whether documents are relevant to the topic, thereby establishing the ground truth. The relevance labels are then used to evaluate ranking algorithms easily~\citep{thakur2021beir}.

The Text REtrieval Conference (TREC) exemplifies community efforts to build large test collections with manual (gold) labels~\citep{voorhees2005robust, roberts2021treccovid, craswell2024deeplearning, upadhyay2024largescalestudyrelevanceassessments}. Although these manual annotations are reliable, a major problem with this process is that they are costly and time-consuming. When the annotation budget is limited, test collections could become too small, potentially leading to unstable evaluation results. Over the years, various document selection techniques have been proposed to use human effort more efficiently~\citep{cormack1998mtf, aslam2006statsmethod, aslam2007inferring, li2017activasampling}, but the problem of data size has always persisted.

To solve the problem of annotation cost, some recent studies have proposed to use Large Language Models (LLMs) to replace the annotation process~\citep{Faggioli2023llmreljudge, thomas2024llmreljudge, qin2024llmranking}. LLM-based assessments can drastically increase the size of test collections since the budget is no longer a problem. Moreover, studies have shown that LLM annotations perform almost always better than other low-cost alternatives like crowd-sourcing~\citep{thomas2024llmreljudge}. However, while these studies have shown some success, we can infer from their results that LLMs do not perfectly match the ground truth judgments. Particularly, LLM judgments are often biased (i.e., either too liberal or too strict~\citep{alaofi2024fooledllm, thomas2024llmreljudge}). Even if a well-performing LLM or prompt is found on one dataset, there is no guarantee that it will perform the same in practice, due to differences in tasks, searchers, topics, or documents. Relying entirely on LLMs may also lead to theoretical challenges like the risk of overfitting to LLM-based metrics~\citep{clarke2024llmbasedrelevanceassessmentcant}. As a result, relying solely on LLMs to build test collections is recently argued to be risky~\citep{soboroff2024dontusellms, clarke2024llmbasedrelevanceassessmentcant}. This tension creates a dilemma: manual judgments are ground truth and high-quality but expensive; LLM judgments are cheap and scalable but error-prone and unreliable.

To address the dilemma between LLMs and humans, our work proposes \textbf{L}LM-\textbf{A}ssisted \textbf{R}elevance \textbf{A}ssessments (\textbf{LARA}) to achieve a reliable, yet budget-friendly annotation procedure. Given a limited budget, LARA uses LLMs to choose the most effective documents for manual annotation. Specifically, LARA uses LLMs' output token probabilities to find their \textit{most uncertain} predictions that should be manually annotated. Then, the rest of the documents (easy and more confident ones) can be judged by LLMs. While directly using LLM relevance probability as \textit{uncertainty} is an effective method to find documents for manual annotation, we argue and show that these LLM predictions are often biased and are not optimal. Therefore, we propose an algorithm that actively learns to calibrate the LLM relevance probabilities, while continuously identifying the most effective documents to annotate manually. The remaining documents are then annotated using the calibrated LLM predictions, adjusted by the calibration model learned from human assessments. Thus, in our algorithm, calibrated LLM predictions guide manual annotations, and the gathered manual annotations in turn refine LLM predictions. Moreover, while most existing algorithms for efficient manual annotations are only suited for binary relevance labels~\citep{cormack1998mtf,aslam2006statsmethod,li2017activasampling}, LARA can deal with the general \textit{graded} relevance framework (e.g., relevance levels 0/1/2), where the binary relevance setting is only a special case of our framework. Finally, we conduct comprehensive experiments on TREC-7 Ad Hoc, TREC-8 Ad Hoc, TREC Robust 2004, and TREC COVID datasets, where we show that our LARA algorithm almost constantly outperforms the tested alternative methods under various budget constraints. We are also the first to study the performance of LLM-based methods and manual methods \textit{under different budget constraints}. 
% Code and scripts for using LARA, as well as reproducing our experiments, will be made publicly available upon the paper's acceptance.

\section{Related Works}
\subsection{Efficient Human Annotations}
When a document collection is large, annotating all topic-document pairs available is difficult due to the immense human labor required. To solve this problem, various methods have been proposed to select subsets of documents for human annotation~\citep{cormack1998mtf, aslam2006statsmethod, aslam2007inferring, li2017activasampling, voorhees2018bandit, ganguly2023queryspecificvariabledepthpooling}.

% \textit{Depth-k pooling} is the most basic and well-used method for building test collections: a set of systems ranks the document collection against each topic, and only the union of the top-\textit{k} retrieved documents from each system for each topic are judged. \textit{Move-to-Front} (MTF)~\citep{cormack1998mtf} is one of the pioneering examples of methods that intelligently select documents to judge that are more likely to be relevant. During each iteration, MTF identifies the highest-scoring unjudged document from the current top-priority system. Then, based on the relevance of the selected document, MTF recalculates the priority of the selected system and proceeds with the iteration. \textit{Multi-armed bandits (MAB)}~\citep{losada2016feelinglucky} is also a popular method to select relevant documents. This method randomly alternates between selecting a document from the best system at the current stage and sampling a document across the entire collection.
Depth-\textit{k} pooling is a common method for building test collections by judging only the union of the top-\textit{k} retrieved documents from each system. \textit{Move-to-Front} (MTF)\citep{cormack1998mtf} adaptively picks unjudged documents and updates system priorities based on relevance, while \textit{multi-armed bandits (MAB)}\citep{losada2016feelinglucky} alternates between the best system at each stage and random sampling. 
While these methods only rely on human annotations, thus providing reliable ground truth judgments, it is costly and time-consuming to build a large test collection. This often results in a small test set and inconsistent evaluations when there is not enough budget.

% In the field of e-Discovery, which is the process of identifying relevant electronic documents, such as for legal proceedings, some active learning methods are used to search for relevant documents~\citep{cormack2104tar, doug2018ediscovery, cormack2016scal, Stevenson2023stopping}. Rahman et al.~\citep{rahman2020activelearningtestcollection} have investigated the effectiveness of these methods on building test collections, where Simple Active Learning (SAL)~\citep{lewis1994uncertaintysampling} and Continuous Active Learning (CAL)~\citep{cormack2009cal} are found to be particularly effective for building test collections. SAL selects documents based on uncertainty sampling, and CAL targets likely-relevant documents. Rahman et al.~\citep{rahman2020activelearningtestcollection} also demonstrated hybrid approaches use the trained model to judge unannotated documents. Although sharing similar concepts with our study, the CAL and SAL hybrid methods can only train their models on the annotated data, while our method guides which documents to annotate based on a calibration model that leverages LLM predictions, making our method more efficient and effective, as we demonstrate in the experiments. 

Active learning methods like Simple Active Learning (SAL)\citep{lewis1994uncertaintysampling} and Continuous Active Learning (CAL)\citep{cormack2009cal}, commonly used in e-Discovery~\citep{cormack2104tar, doug2018ediscovery, cormack2016scal, Stevenson2023stopping}, have also been applied to building test collections~\citep{rahman2020activelearningtestcollection}. SAL selects documents based on uncertainty sampling, while CAL focuses on likely-relevant documents. Rahman et al.~\citep{rahman2020activelearningtestcollection} also demonstrate hybrid approaches that use the trained model to judge unannotated documents. Although these hybrid methods share similar concepts with our study, our method employs a calibration model that leverages LLM predictions to guide annotation, achieving greater efficiency and effectiveness, as we show in our experiments.

\subsection{Automated Relevance Assessments}
Due to the high cost and effort needed for building test collections with ground truth labels, some have previously considered using cheaper third-party assessors~\citep{alonso2011crowd,eickhoff2012crowd}. However, working with these third-party labelers, especially crowd workers, can often suffer from misunderstanding the searcher needs, mistakes, collusion, ``spammy'' workers, and other biases~\citep{clough2013crowdsource,thomas2022crowd}.

Researchers have also investigated fully automatic methods to create test collections~\citep{dietz2022automaticwiki, Dietz2020autotestcollection}. Particularly, an increasing number of researchers are exploring the ability of LLMs in relevance judgments~\citep{Faggioli2023llmreljudge, thomas2024llmreljudge, MacAvaney2023llmreljudge, upadhyay2024llmspatchmissingrelevance, rahmani2024llmfortestcollections, abbasiantaeb2024uselargelanguagemodels, mehrdad2024largelanguagemodelsrelevance, alaofi2024fooledllm, upadhyay2024largescalestudyrelevanceassessments}. 

These studies have shown the capability of LLMs in relevance assessments and have found that LLM judgements somewhat correlate with the ground truth labels and often perform better than crowd workers~\citep{thomas2024llmreljudge}. However, the relevance predictions by LLMs are often biased. For example, Alaofi et al.~\citep{alaofi2024fooledllm} report that LLMs tend to over-generate relevance labels compared to the ground truth. If we were to build a test collection only using LLMs, we would not be able to avoid such bias, making a full replacement risky~\citep{soboroff2024dontusellms, clarke2024llmbasedrelevanceassessmentcant}. Even if a well-performing prompt is found on a certain dataset, there is no guarantee that it will work on an unseen task or data that we want to use for a test collection. To minimize such risks posed by LLMs, our LARA algorithm effectively uses partial ground truth judgments collected under a limited budget to actively calibrate the LLM predictions, building a more reliable dataset. Fine-tuned LLMs have shown improvement in accuracy~\citep{meng2024queryperformancepredictionusing}, but have crucial limitations in annotation latency, as we will discuss in Section~\ref{sec: discussions}.

Upadhyay et al.~\citep{upadhyay2024llmspatchmissingrelevance} and Abbasiantaeb et al.~\citep{abbasiantaeb2024uselargelanguagemodels} report the effect of using LLMs to fill relevance judgment holes that humans could not judge. Unlike these \textit{post-hoc }procedures, our online approach leverages LLM predictions to \textit{guide} which documents to manually judge, continually refining both the LLM calibration and our selection strategy.
% In contrast, we consider an online procedure using an LLM to \textit{guide} which documents to manually judge.

\section{LLM-Assisted Relevance Assessments}
\begin{figure*}[tp]
    \vspace{-10pt}
    \begin{tabular}{cc}
        \includegraphics[width=2.90in, keepaspectratio]{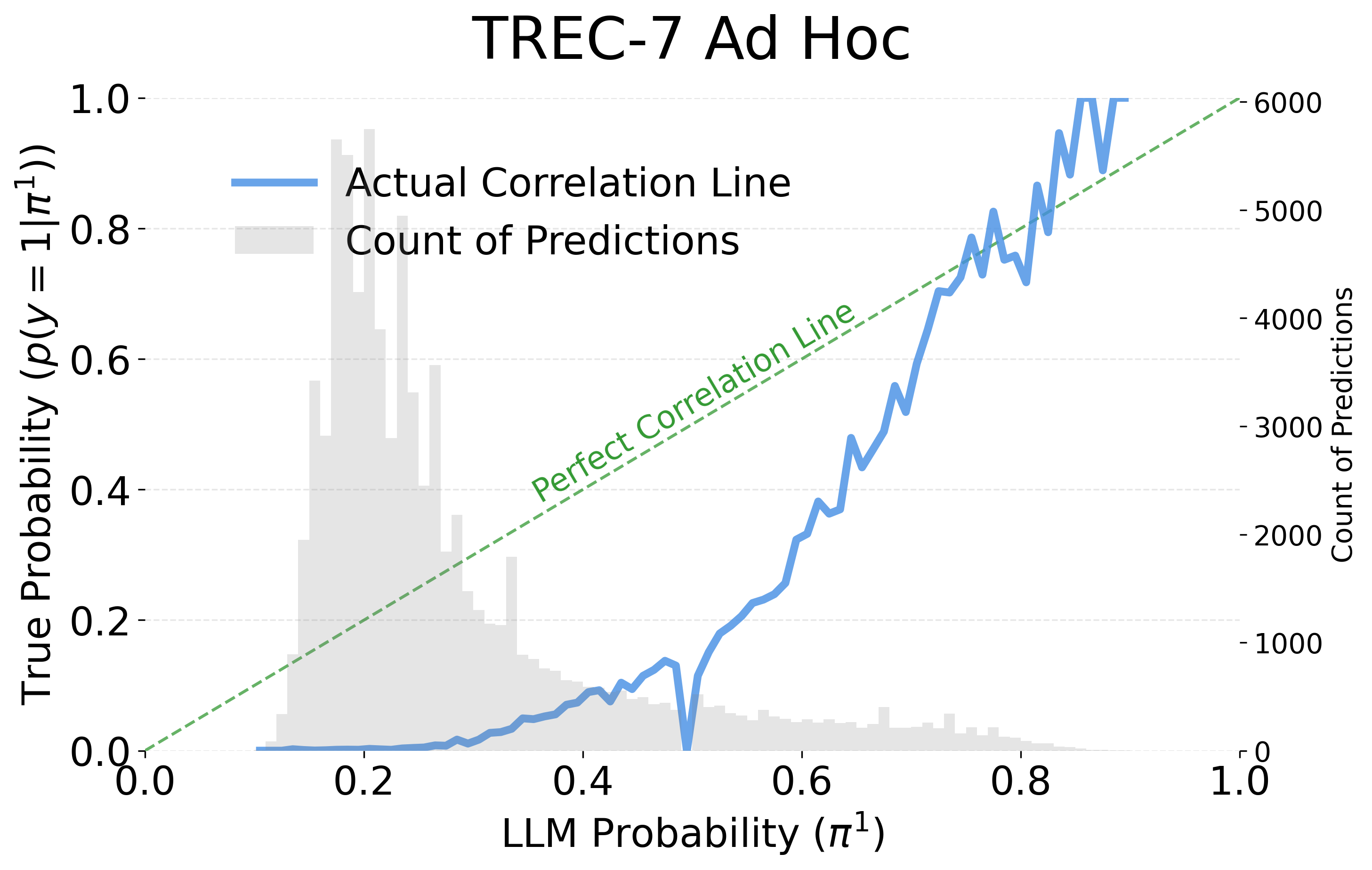} &
        \includegraphics[width=2.90in, keepaspectratio]{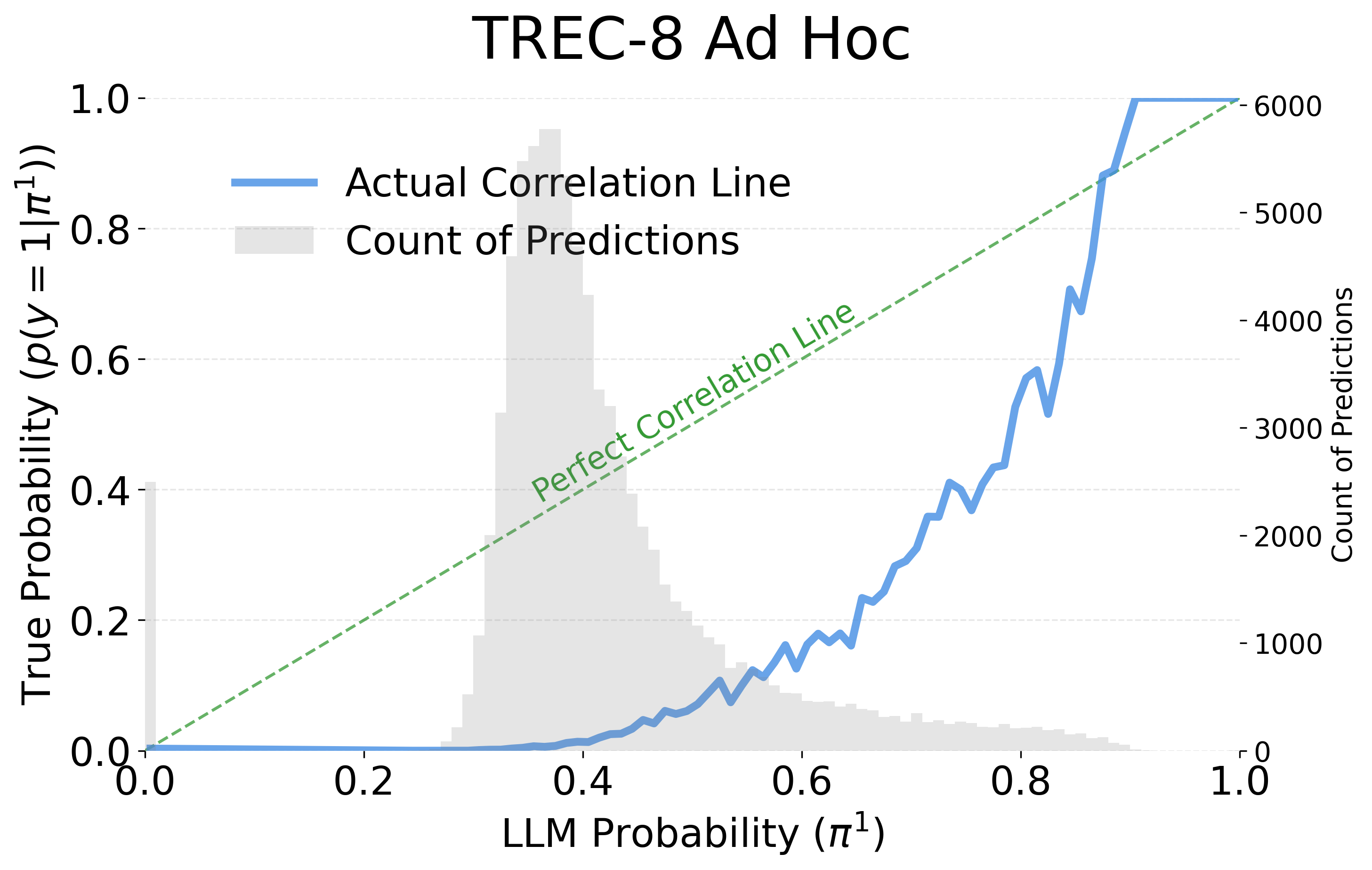} \\ [3pt]
        \includegraphics[width=2.90in, keepaspectratio]{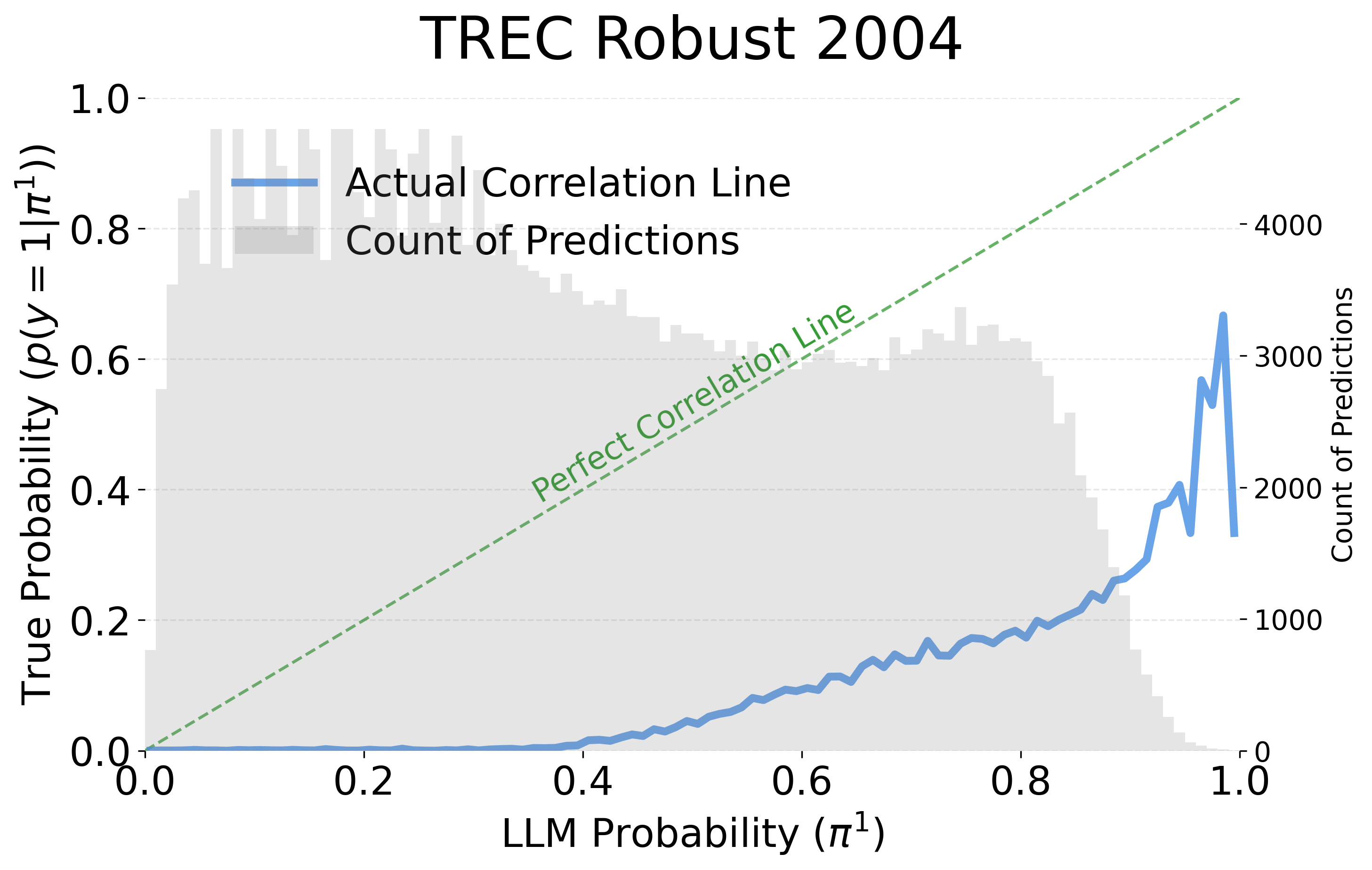} &
        \includegraphics[width=2.90in, keepaspectratio]{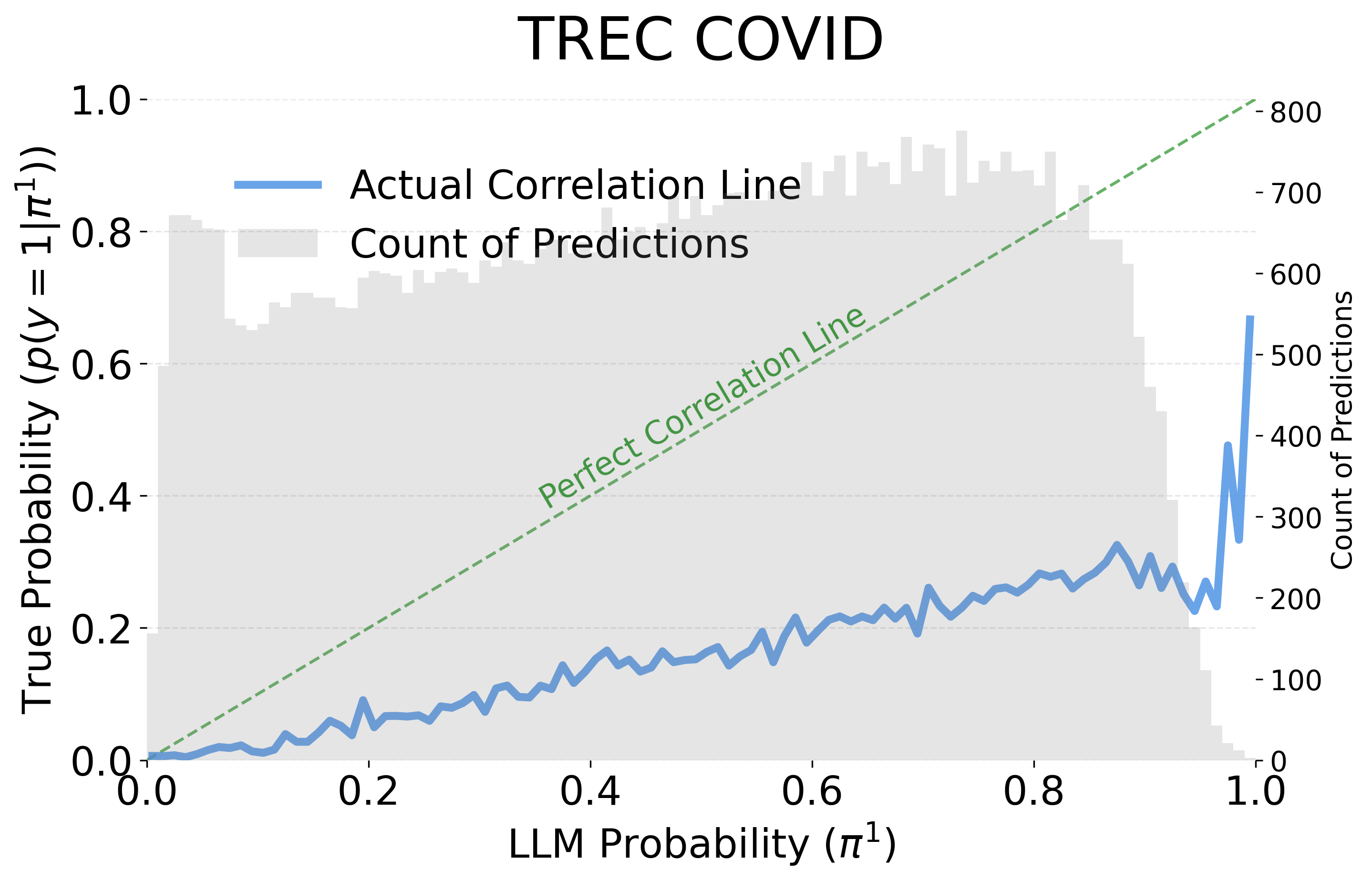}
    \end{tabular}
    \vspace{-12pt}
    \Description{Four scatter–histogram panels comparing LLM-predicted 
    relevance probabilities with true relevance on four TREC test 
    collections.  In every panel, grey bars show the distribution of LLM 
    outputs, a blue line plots the empirical true-relevance rate, and a 
    dotted 45-degree line marks perfect calibration; the blue line falls 
    well below the dotted line except near probability 1.0, indicating the 
    LLM is generally over-confident.}
    \caption{Showing the ground truth relevance probability ($p(y=1|\pi^1)$) at each LLM predicted relevance probability ($\pi^1$) on TREC-7 Ad Hoc (top left), TREC-8 Ad Hoc (top right), TREC Robust04 (bottom left), and TREC-COVID (bottom right). The green line shows the correlation line if the correlation is perfect. For reference, the gray bars (secondary y-axis) show the number of predictions at each LLM probability value ($\pi^1$), showing where the LLM predictions are most concentrated. 
    % TREC-7 Ad Hoc (top left) and TREC-8 Ad Hoc (top right) are test collections with binary relevance (i.e., $y=0/1$), and TREC Robust04 (bottom left) and TREC-COVID (bottom right) consist of three-level relevance grades (i.e., $y=0/1/2$).
    }
    \label{fig:correlations}
\vspace{-10pt}
\end{figure*}
% In this section, we examine how to build a large and reliable test collection using both human and LLM judgments. 
% Once we introduce the problem setup, we start by proposing a naive approach. We then demonstrate, both theoretically and empirically, why this naive approach can be sub-optimal and derive our LARA algorithm based on this understanding. 

% In this section, we examine how to build a large and reliable test collection using both human and LLM judgments. Once we introduce the problem setup, we start by proposing the use of LLM relevance probabilities and a naive approach that straightforwardly uses them. We then demonstrate, both theoretically and empirically, why this naive approach can be sub-optimal. Building on this understanding, we derive our LLM-Assisted Relevance Assessments (LARA) algorithm. We consider the general graded relevance setting, where the binary relevance setting (i.e., relevant/non-relevant) can be interpreted as a special case of our general framework.

\subsection{Problem Setup}
Let $D=\{x_1, x_2, ..., x_{|D|}\}$ be the full dataset to annotate, where $x_i$ is the $i$-th data point (i.e., topic-document pair). We consider the graded relevance setting where each data point has an associated (ground-truth) human annotation of $y\in\{0, 1, ..., l\}$, representing a graded relevance label of $l+1$ levels. The binary relevance setting is a special case of our framework. We are given an annotation budget $B (\leq |D|)$, which is the maximum number of ground truth annotations that the human assessors can perform. The goal is to build a test collection that could accurately rank the evaluated IR systems in the order of performance.

While humans (topic developers) can provide the ground truth relevance labels, obtaining a sufficiently large fully human-annotated test collection can be infeasible due to the limited budget $B$. As a potential solution, recent studies suggest using LLM predictions to generate synthetic relevance assessments at scale \citep{Faggioli2023llmreljudge, thomas2024llmreljudge}. However, while LLM annotations are cheap, they are not perfectly accurate and are unreliable~\citep{alaofi2024fooledllm, thomas2024llmreljudge}. This imbalance motivates our hybrid strategy: use human effort for the most effective annotations while relying on (calibrated) LLM assessments for the bulk of the data.

\label{sec: naive}

\subsection{A Naive Uncertainty-Based Approach}
Ideally, to minimize the number of annotation errors in the test collection, we want to spend our manual annotation budget on data points where LLMs are wrong. However, in practice, we cannot know this until the manual annotation has taken place. Thus, a straightforward strategy could be to focus manual effort on data points where the LLM is \textit{uncertain}. Intuitively, if the LLM is unsure of a particular label, verifying that prediction would seem more effective than annotating data points where the LLM is confident. 

To implement this idea, we propose to use the normalized LLM prediction probability $\pi^k=\pi(y=k|x)$ of each relevance level $k$. The probability $\pi^k$ is calculated for each data point $x_i$ as
\begin{equation}
    \pi^k_i=\pi(y=k|x_i)=\frac{p_\text{LLM}(k|x_i)}{\sum_{j=0}^lp_\text{LLM}(j|x_i)},
\end{equation}
where $p_\text{LLM}(k|x)$ is the probability of the LLM assigning the relevance level $k$ to a document given a query. We express the LLM probability set for data point $x_i$ as $\pi_i=\{\pi_i^0, \pi_i^1, ..., \pi_i^l\}$.

Given these probabilities, one naive approach is to identify the data points on which LLM is the most uncertain and annotate those manually, which we call the \textbf{Naive} method. A simple heuristic is to first determine the most likely predicted level $k'$ and the second most likely predicted level $s'$ for each data point
\begin{equation}
    k'_i = \argmax_j \pi_i^j \quad \text{and} \quad s'_i = \argmax_{j \neq k'_i} \pi_i^j.
\end{equation}
Then, the LLM's most uncertain data point $x_i$ would be where the margin $m'_i=\pi_i^{k'}-\pi^{s'}_i$ is the smallest. Thus, the Naive method manually annotates $B$ data points with the smallest margins $m'$, and the remaining data points are given the LLM predictions $\hat{y}=k'$.

\subsection{LLM-Assisted Relevance Assessments}
\label{sec:binary}
% Although it may seem like the Naive method would work well, it has a critical flaw: it assumes that the LLM’s predicted probabilities align closely with the true relevance probabilities. In reality, LLM predictions can be systematically biased or poorly calibrated. For instance, the LLM might consistently overestimate relevance for certain types of documents or under-detect relevance for others \citep{thomas2024llmreljudge}. As a result, selecting documents based solely on the LLM’s raw probabilities may not pinpoint where the true uncertainty lies.

Although it may seem like the Naive method would work well, LLMs' uncertain data points are not necessarily the best data points to annotate, mainly because LLM predictions are often biased~\citep{thomas2024llmreljudge}. To show this and to develop a better method, we first introduce a theorem that shows the optimal inspection point for such a classification problem, built on the concept of uncertainty sampling.
\begin{theorem}[Optimal Inspection Point\footnote{Proof can be found at \url{https://github.com/RikiyaT/LARA/blob/main/docs/proof.md}}]
    Let \(p(y=j | \pi)\) denote the probability of the true relevance level being \(j\) given LLM probabilities \(\pi\). Let \(k = \argmax_{j} p(y=j | \pi)\) be the level with the highest probability and \(s = \argmax_{j\neq k} p(y=j | \pi)\) be the level with the second highest probability. Then, the greatest potential reduction in classification error is achieved by annotating the data point with the smallest margin
    \begin{equation*}
        m=p(y=k|\pi)-p(y=s|\pi),
    \end{equation*}
    where the remaining data are classified into $\hat{y}=\text{argmax}_jp(y=j|\pi)$.
\label{theo:opt}
\end{theorem}
Theorem~\ref{theo:opt} indicates that the data point with the smallest \textit{naive} margin $m'$ may not be the best point to annotate unless $\pi$ is perfectly correlated with the true relevance probabilities $p(y=j|\pi^j)$. To empirically show some examples of the correlation, Figure~\ref{fig:correlations} shows the relationship between the Llama-3.1-70B-Instruct predictions $\pi^1$ and the true relevance probabilities $p(y=1|\pi^1)$, on the TREC-7 Ad Hoc, TREC-8 Ad Hoc, TREC Robust04, and the TREC COVID dataset. Refer to Table~\ref{tab:datasets} for statistics of each dataset. From the figure, we see that the correlation is far from perfect alignment, which shows that selecting documents based solely on the LLM’s raw probabilities may not pinpoint where the true uncertainty lies.

From Theorem~\ref{theo:opt}, we know that we should instead annotate data points with the smallest true margin $m$. However, since the true probabilities $p(y=j|\pi)$ are unknown, we propose to learn this calibration mapping for each relevance level $j\in \{0, ..., l\}$ while selecting reasonable and effective data points for manual annotation.

Now, we explain our \textbf{LARA} algorithm. A more detailed explanation is provided in Algorithm~\ref{alg:lara}. First, the calibration is set to $\hat{p}(y|\pi)=\pi$. Thus, we start by manually annotating data points closest to $m'=\pi_i^{k'}-\pi^{s'}_i$ like the Naive method. Then, using these initial manual annotations, we train a calibration estimator and gradually update the prediction $\hat{p}(y|\pi)$ for all relevance levels. The calibration can be performed by regressing the true relevance $y$ on the estimated relevance probability $\pi$. We then repeat the process of annotating data points with the smallest predicted margin
\begin{equation*}
    \hat{m}=\hat{p}(y=k|\pi)-\hat{p}(y=s|\pi),
\end{equation*} where $k=\argmax_{j}\hat{p}(y=j|\pi)$ and $s=\argmax_{j\neq k}\hat{p}(y=j|\pi)$, and we iteratively update the calibration until budget $B$ is exhausted. Finally, for the data points that were not manually annotated, we assign $\hat{y}=\argmax_{j}\hat{p}(y=j|\pi)$. This final step allows the data points that were not manually annotated to be given calibrated and debiased LLM predictions learned on limited ground-truth labels.

\begin{algorithm}[tp]
\caption{LARA: LLM-Assisted Relevance Assessments}
\label{alg:lara}
\begin{algorithmic}[1]
\State \textbf{Input:} Dataset $D=\{x_1, x_2, ..., x_{|D|}\}$, LLM predicted probability vectors $\{\pi_i\}_{i=1}^{|D|}$ where $\pi_i=\{\pi_i^0, \pi_i^1, \ldots, \pi_i^l\}$, budget $B$
\State \textbf{Output:} Truth labels $y_A$ and predicted annotations $\hat{y}_{D\setminus A}$

\State Initialize $\hat{p}(y=j|\pi)=\pi^j$ for all $j\in\{0,...,l\}$ and for all $x_i \in D$
\State $A \leftarrow \emptyset$ 
\While{$|A| < B$}
    \For{$x_i \in D \setminus A$}
        \State $k = \argmax_{j} \hat{p}(y=j|\pi_i)$
        \State $s = \argmax_{j\neq k}\hat{p}(y=j|\pi_i)$
        \State Compute the margin $\hat{m}_i = \hat{p}(y=k|\pi_i) - \hat{p}(y=s|\pi_i)$
    \EndFor
    \State $x^* \leftarrow \arg\min_{x_i \in D \setminus A} \hat{m}_i$ 
    \State $y^* \leftarrow \text{ManualAnnotation}(x^*)$
    \State $A \leftarrow A \cup \{(x^*, y^*)\}$
    \State UpdateCalibrator$(\hat{p}, A)$ 
\EndWhile

\State $y_A \leftarrow \{y^* : (x^*, y^*) \in A\}$
\For{$x_i \in D \setminus A$}
    \State $\hat{y}_i = \argmax_{j}\hat{p}(y=j|\pi_i)$
\EndFor
\State \Return $(y_A, \hat{y}_{D\setminus A})$
\end{algorithmic}
\end{algorithm}

\subsection{Accessibility for Real-World Implementation}
\label{sec:num_annot}
While our LARA algorithm appears highly effective, it may lack practicality. This is because, in the proposed algorithm, LARA is not suited for having multiple (gold) assessors. In venues that build these test collections, coordinators usually assign multiple assessors to come up with topics and assess documents~\citep{voorhees2000overview}. Usually, each assessor comes up with a number of their own topics based on their own interests and then judges the relevance of the given documents. However, if the LARA algorithm is used by multiple assessors, each assessor would have to assess documents whenever their topic is chosen; this requires all assessors to be working at the same time.

To solve this problem, we bundle the topics into groups, based on the number of assessors $n$. Then, we divide the budget $B$ by the number of assessors; each group of topics would be assigned with the budget $\frac{B}{n}$. We only sample from one group until the budget $\frac{B}{n}$ is exhausted, then we move on to the next group. This way, assessors do not have to be working at the same time. In the experiments, we denote the number of assessors as $\text{LARA}(n)$. For example, we write $\mathrm{LARA}(n=3)$ if there are 3 annotators. When each topic has a different annotator, we write $\mathrm{LARA}(n=N)$ where $N$ is the total number of topics. We test the grouped methods in our experiments to see how the performance of LARA changes with different numbers of annotators.

% Some works focus on finding as many relevant documents as possible~\citep{cormack1998mtf, losada2016feelinglucky}, but this process is recently argued to be unfair~\citep{li2017fair, voorhees2018bandit}. 

\section{Experiment Setup}
\label{sec: data}
In this section, we conduct experiments to see the performance of LARA compared to the alternative methods of building test collections under different budget constraints $B$. In the experiments, the budget is shown as a ratio against the total number of assessments in the full test collection (i.e., $B/|D|$). The statistics of the datasets are provided in Table~\ref{tab:datasets}. All datasets are originally constructed via the Depth-$k$ method. Each dataset contains runs submitted by the participating systems, where each run gives a ranking of documents given a topic. We aim to build a test collection that can accurately rank these systems in the order of performance.

\begin{table}[h]
\centering
\vspace{-5pt}
\caption{Statistics of the datasets used in our experiments.}
\vspace{-8pt}
\label{tab:datasets}
\begin{tabular}{lcrr}
\toprule
\textbf{Dataset} & \textbf{Relevance Levels} & \textbf{\#Data Points} & \textbf{\#Systems} \\
\midrule
TREC-7 Ad Hoc     & 0, 1                   & \raggedleft 80,346      & 103 \\
TREC-8 Ad Hoc     & 0, 1                   & \raggedleft 86,829      & 131 \\
TREC Robust04     & 0, 1, 2               & \raggedleft 311,392     & 110 \\
TREC-COVID        & 0, 1, 2               & \raggedleft 60,075      & 126 \\
\bottomrule
\vspace{-15pt}
\end{tabular}
\end{table}

\subsection{Evaluation Metrics}
For each system performance evaluation, we use the \textit{Mean Average Precision} (MAP) score for the binary relevance experiments and the \textit{Normalized Discounted Cumulative Gain} (NDCG) score for the graded relevance experiments. Then, based on the MAP/NDCG scores, we rank the participating systems. We consider the system rankings evaluated by the full test collection to be the ground truth ranking. We evaluate the system rankings created under budget by computing Kendall's $\tau$, which will be our main metric to evaluate the method's performance. If the evaluated ranking is identical to the ground truth ranking built by the full test collection, the score will be 1, which would mean that it is a perfect ranking.

In addition to the main performance evaluation through Kendall's $\tau$, we report the Maximum Drop, which shows the maximum decrease in rank for a system in the ground truth ranking compared to the evaluated ranking. The Maximum Drop has been reported in recent studies because Kendall's $\tau$ scores over large lists can indicate general concordance despite a few items with large rank differences~\citep{voorhees2018bandit, voorhees2022relevants}. Higher Maximum Drop values indicate that at least one system is being unfairly treated in the ranking.

One crucial advantage of our algorithm, beyond selecting effective documents to manually annotate, is that they can minimize the errors of the LLM annotations. Specifically, LARA calibrates the LLM predictions by learning a model $\hat{p}(y|\pi)$ from the collected human annotations. Moreover, in LARA (and in the Naive method), we try to give ground-truth labels to the most uncertain data points, which also helps minimize the LLM prediction errors. Thus, as exploratory analyses, we show the performance increase of LLM annotations with LARA by evaluating via \textit{overlap} score~\citep{voorhees1998overlap}.
\begin{equation}
    \text{Overlap} = \frac{\text{\# of True Positives}}{\text{\# of True Positives}+\text{\# of False Predictions}}\,\,\,\,\, ,
\end{equation}
where True Positives are when ground truth and LLM both judge the same relevance and are not non-relevant (i.e., $y_i=\hat{y}_i\geq 1$), and the False Predictions are all the false predictions (i.e., $y_i\neq \hat{y}_i$). We use the overlap metric over Cohen's Kappa because the number of non-relevant labels greatly exceeds the number of relevant documents in the experimented test collections.

\subsection{Comparison Methods} 
We compare the LARA algorithm with several baseline approaches. We first introduce methods from previous studies that only use manual assessments. Note that since most of these existing manual methods are only suited for binary relevance levels, we binarize the relevance levels in the learning procedure by setting a threshold at $\frac{l}{2}$, but we keep the graded relevance levels in their final annotations.

\textbf{Depth-k Pooling.} Depth-\textit{k} Pooling is a method with only human assessments. It assesses the top-\textit{k} documents at each rank for each topic for all systems until the budget is empty.

\textbf{Move-to-Front Pooling (MTF)~\citep{cormack1998mtf}.} In the MTF algorithm, all systems start with equal priority. On each round, it selects the first unjudged document provided by the highest-priority system to judge. If relevant, it continues; if not, the system's priority is lowered, and the next system is selected.

\textbf{Multi-armed Bandits (MAB)~\citep{losada2016feelinglucky}.} This method treats selection as a multi-armed bandit problem, randomly choosing between the top-priority system or sampling from the entire collection. For the MAB baseline, we used the best-performing method, \textbf{MaxMean Non-Stationary (MM-NS)}, with default settings from \citep{losada2016feelinglucky}.

\textbf{Continuous Active Learning (CAL)~\citep{cormack2009cal}.} CAL selects the most likely relevant data point to annotate and continuously trains a regression model. We compare \textbf{CAL(human)}, which consists of only manual annotations, and \textbf{CAL(hybrid)}, which also predicts the remaining data points with the trained regression model.

\textbf{Simple Active Learning (SAL)~\citep{lewis1994uncertaintysampling}.} Similar to CAL, SAL iteratively trains a regression model, but SAL instead selects data points based on uncertainty sampling. We compare both \textbf{SAL(human)} and \textbf{SAL(hybrid)} in the experiments.

In addition to the above manual methods, we add three more baselines that incorporate LLM predictions:

\textbf{LLM-Only.} This method uses only LLM judgments to build a test collection.

\textbf{Random.} This method uses both LLM annotations and human annotations. $B$ randomly chosen data points are manually annotated and the rest are judged by LLMs. This method can also be seen as using LLMs to \textit{fill the holes} of manual judgments~\citep{upadhyay2024llmspatchmissingrelevance, abbasiantaeb2024uselargelanguagemodels}.

\textbf{Naive.} This method uses both humans and LLMs to build a test collection. It chooses the LLM's most uncertain data points to manually annotate by naively using LLM relevance probabilities $\pi$, then annotates the rest using LLM predictions. Further details of this method are provided in Section~\ref{sec: naive}. Note that the Naive method was never introduced by any of the previous studies; we compare this method as a baseline only to visualize the advantage of our LARA calibration algorithm.

We investigate the effect of having different numbers of annotators. Thus, we show results of \textbf{LARA($\boldsymbol{n=1}$)}, \textbf{LARA($\boldsymbol{n=3}$)}, and \textbf{LARA($\boldsymbol{n=N}$)}. More on the number of annotators is in Section~\ref{sec:num_annot}. Logistic regression is used as the calibration model for LARA.

To study the effect of the prompt on LARA, we test three different prompts\footnote{Prompts are available at \url{https://github.com/RikiyaT/LARA\#-prompt-templates}}. In addition to our simplest base prompt, we also share the results with the \textbf{rationale prompt} based on Upadhyay et al.~\citep{upadhyay2024largescalestudyrelevanceassessments} and the \textbf{utility prompt} based on Thomas et al.~\citep{thomas2024llmreljudge} as an extensive study. The prompts are slightly modified from the original to better align with our task. Methods that use these prompts are named in the format \textbf{MethodName(PromptType)}, like \textbf{LARA(rationale)} and \textbf{LARA(utility)}, where the LARA methods here use LARA($n=N$) for the number of annotators.

Unless specified otherwise, all LLM-based baseline methods use Llama-3.1-70B-Instruct\footnote{https://ai.meta.com/blog/meta-llama-3-1/}. Temperature is set to $3.0$, and top-$k$ is set to 50, so the LLM can output the probabilities of answering various relevance grades. To study the effect of LLM performance on LARA, we also show results using the lightweight variant of the base LLM, Llama-3.1-8B-Instruct, as an ablation study. Methods that use the 8B-version are named \textbf{LLM-Only (Llama-8B)}, \textbf{Random (Llama-8B)}, \textbf{Naive (Llama-8B)}, and \textbf{LARA (Llama-8B)}, where LARA (Llama-8B) uses LARA($n=N$) for the number of annotators.

\begin{table*}[t]
\centering
\caption{Kendall's Tau correlation scores (larger is better) for the ranked systems based on MAP scores across different methods and different ratios of manual annotation budget. The first blocks show the main results, where for each column, the top \textit{three} highest scores are shown in bold, the fourth is in italics, and the fifth is underlined. For the methods based on the rationale and utility prompt, only the top score \textit{of each prompt} is shown in bold. For the methods based on Llama-8B, the top score is highlighted in bold. Values in parentheses show the Maximum Drop in the system rank (smaller is better).}
\centering
\small
\vspace{-10pt}
\begin{tabular}{l|rrrrrrrrrr}
\hline
% =========================
%  TREC-7 Ad Hoc
% =========================
\multicolumn{11}{c}{\textbf{TREC-7 Ad Hoc}} \\
\hline
&\multicolumn{10}{c}{\textbf{Ratio of Manual Annotation Budget ($B/|D|$)}} \\
\textbf{Method} & \textbf{$\frac{1}{512}$} & \textbf{$\frac{1}{256}$} & \textbf{$\frac{1}{128}$} & \textbf{$\frac{1}{64}$} & \textbf{$\frac{1}{32}$} & \textbf{$\frac{1}{16}$} & \textbf{$\frac{1}{8}$} & \textbf{$\frac{1}{4}$} & \textbf{$\frac{1}{2}$} & all \\
\hline
Depth-k & 0.495 (65) & 0.624 (83) & 0.704 (72) & 0.751 (47) & 0.838 (39) & \underline{0.879} (18) & \textit{0.922} (13) & \textit{0.931} (5) & \textbf{0.973} (3) & \textbf{1.000} (0) \\
MTF & 0.508 (58) & 0.578 (56) & 0.657 (56) & 0.694 (56) & 0.675 (45) & 0.764 (39) & 0.810 (28) & 0.837 (24) & 0.920 (15) & \textbf{1.000} (0) \\
MM-NS & 0.405 (56) & 0.405 (56) & 0.465 (61) & 0.747 (42) & 0.802 (32) & 0.813 (25) & 0.874 (18) & 0.923 (8) & \underline{0.961} (4) & \textbf{1.000} (0) \\
CAL(human) & 0.154 (59) & 0.225 (65) & 0.212 (74) & 0.284 (73) & 0.674 (63) & 0.650 (39) & 0.822 (25) & 0.839 (33) & 0.897 (17) & \textbf{1.000} (0) \\
CAL(hybrid) & 0.459 (71) & 0.414 (74) & 0.372 (71) & 0.509 (85) & 0.583 (50) & 0.693 (26) & 0.788 (27) & 0.792 (52) & 0.895 (21) & \textbf{1.000} (0) \\
SAL(human) & 0.471 (63) & 0.493 (33) & 0.460 (62) & 0.534 (61) & 0.660 (84) & 0.784 (29) & 0.786 (29) & 0.830 (23) & 0.887 (15) & \textbf{1.000} (0) \\
SAL(hybrid) & 0.317 (60) & 0.467 (60) & 0.342 (84) & 0.347 (78) & 0.324 (38) & 0.784 (38) & 0.827 (24) & 0.843 (28) & 0.899 (14) & \textbf{1.000} (0) \\
LLM-Only & \underline{0.861} (42) & \underline{0.861} (42) & \underline{0.861} (42) & \underline{0.861} (42) & 0.861 (42) & 0.861 (42) & 0.861 (42) & 0.861 (42) & 0.861 (42) & 0.861 (42) \\
Random & \textit{0.862} (42) & 0.861 (42) & 0.861 (42) & 0.861 (42) & \underline{0.864} (38) & 0.865 (38) & 0.873 (37) & 0.885 (34) & 0.911 (24) & \textbf{1.000} (0) \\
Naive & 0.861 (42) & \textit{0.862} (41) & \textit{0.863} (38) & \textit{0.864} (38) & \textit{0.868} (37) & \textit{0.883} (36) & \underline{0.905} (30) & \underline{0.923} (23) & 0.938 (15) & \textbf{1.000} (0) \\
\textbf{LARA($n=1$)} & \textbf{0.869} (36) & \textbf{0.877} (36) & \textbf{0.894} (31) & \textbf{0.900} (30) & \textbf{0.910} (27) & \textbf{0.917} (23) & \textbf{0.930} (20) & \textbf{0.946} (10) & \textbf{0.973} (6) & \textbf{1.000} (0) \\
\textbf{LARA($n=3$)} & \textbf{0.875} (36) & \textbf{0.882} (36) & \textbf{0.894} (31) & \textbf{0.900} (30) & \textbf{0.907} (27) & \textbf{0.916} (23) & \textbf{0.929} (20) & \textbf{0.942} (12) & \textbf{0.980} (4) & \textbf{1.000} (0) \\
\textbf{LARA($n=N$)} & \textbf{0.875} (36) & \textbf{0.884} (36) & \textbf{0.893} (32) & \textbf{0.903} (29) & \textbf{0.912} (28) & \textbf{0.922} (23) & \textbf{0.936} (19) & \textbf{0.941} (11) & \textit{0.966} (7) & \textbf{1.000} (0) \\
\cline{1-11}
LLM-Only(rationale) & 0.837 (42) & 0.837 (42) & 0.837 (42) & 0.837 (42) & 0.837 (42) & 0.837 (42) & 0.837 (42) & 0.837 (42) & 0.837 (42) & 0.837 (42) \\
Random(rationale) & 0.837 (42) & 0.837 (42) & 0.840 (42) & 0.839 (42) & 0.842 (41) & 0.842 (41) & 0.852 (38) & 0.865 (35) & 0.906 (24) & \textbf{1.000} (0) \\
Naive(rationale) & 0.839 (42) & 0.840 (42) & 0.842 (41) & 0.845 (41) & 0.858 (39) & 0.872 (36) & 0.899 (30) & 0.929 (22) & 0.950 (16) & \textbf{1.000} (0) \\
\textbf{LARA(rationale)} & \textbf{0.846} (41) & \textbf{0.857} (36) & \textbf{0.866} (35) & \textbf{0.870} (34) & \textbf{0.881} (31) & \textbf{0.915} (21) & \textbf{0.928} (21) & \textbf{0.942} (15) & \textbf{0.982} (5) & \textbf{1.000} (0) \\
LLM-Only(utility) & 0.846 (40) & 0.846 (40) & 0.846 (40) & 0.846 (40) & 0.846 (40) & 0.846 (40) & 0.846 (40) & 0.846 (40) & 0.846 (40) & 0.846 (40) \\
Random(utility) & 0.846 (40) & 0.846 (40) & 0.847 (40) & 0.847 (40) & 0.848 (39) & 0.847 (37) & 0.857 (36) & 0.873 (33) & 0.912 (22) & \textbf{1.000} (0) \\
Naive(utility) & 0.845 (40) & 0.845 (40) & 0.847 (39) & 0.853 (39) & 0.866 (37) & 0.881 (31) & 0.905 (20) & 0.926 (19) & 0.948 (13) & \textbf{1.000} (0) \\
\textbf{LARA(utility)} & \textbf{0.863} (35) & \textbf{0.873} (33) & \textbf{0.880} (30) & \textbf{0.891} (27) & \textbf{0.892} (24) & \textbf{0.908} (20) & \textbf{0.926} (17) & \textbf{0.940} (12) & \textbf{0.971} (7) & \textbf{1.000} (0) \\
\cline{1-11}
 LLM-Only (Llama-8B) & 0.578 (93) & 0.578 (93) & 0.578 (93) & 0.578 (93) & 0.578 (93) & 0.578 (93) & 0.578 (93) & 0.578 (93) & 0.578 (93) & 0.578 (93) \\
Random (Llama-8B) & 0.579 (93) & 0.577 (93) & 0.579 (93) & 0.579 (93) & 0.580 (93) & 0.581 (93) & 0.595 (92) & 0.621 (88) & 0.696 (77) & \textbf{1.000} (0) \\
Naive (Llama-8B) & 0.578 (93) & 0.579 (93) & 0.579 (93) & 0.581 (93) & 0.585 (92) & 0.595 (92) & 0.605 (92) & 0.643 (85) & 0.730 (62) & \textbf{1.000} (0) \\
\textbf{LARA (Llama-8B)} & \textbf{0.643} (85) & \textbf{0.676} (82) & \textbf{0.715} (72) & \textbf{0.737} (61) & \textbf{0.746} (58) & \textbf{0.756} (52) & \textbf{0.790} (41) & \textbf{0.859} (32) & \textbf{0.924} (15) & \textbf{1.000} (0) \\
\hline
% =========================
%  TREC-8 Ad Hoc
% =========================
\multicolumn{11}{c}{\textbf{TREC-8 Ad Hoc}} \\
\hline
& \multicolumn{10}{c}{\textbf{Ratio of Manual Annotation Budget ($B/|D|$)}} \\
\textbf{Method} & \textbf{$\frac{1}{512}$} & \textbf{$\frac{1}{256}$} & \textbf{$\frac{1}{128}$} & \textbf{$\frac{1}{64}$} & \textbf{$\frac{1}{32}$} & \textbf{$\frac{1}{16}$} & \textbf{$\frac{1}{8}$} & \textbf{$\frac{1}{4}$} & \textbf{$\frac{1}{2}$} & all \\
\hline
Depth-k           & 0.547 (95) & 0.597 (74) & 0.657 (99) & 0.699 (88) & 0.790 (62) & \textit{0.882} (22) & \textit{0.913} (19) & \textbf{0.940} (12) & \textbf{0.982} (6) & \textbf{1.000} (0) \\
MTF               & 0.291 (90) & 0.281 (90) & 0.320 (97) & 0.340 (99) & 0.323 (97) & 0.590 (93) & 0.758 (43) & 0.830 (28) & 0.915 (18) & \textbf{1.000} (0) \\
MM-NS             & 0.319 (98) & 0.314 (100) & 0.478 (76) & 0.525 (92) & 0.726 (42) & 0.756 (45) & 0.844 (43) & \underline{0.888} (23) & \underline{0.957} (13) & \textbf{1.000} (0) \\
CAL(human)        & 0.459 (93) & 0.560 (65) & 0.509 (112) & 0.547 (65) & 0.505 (60) & 0.671 (54) & 0.722 (36) & 0.773 (37) & 0.891 (28) & \textbf{1.000} (0) \\
CAL(hybrid)       & 0.590 (81) & 0.627 (107) & 0.494 (96) & 0.627 (43) & 0.535 (45) & 0.688 (68) & 0.726 (54) & 0.754 (47) & 0.889 (29) & \textbf{1.000} (0) \\
SAL(human)        & 0.325 (99) & 0.328 (68) & 0.620 (88) & 0.522 (65) & 0.574 (70) & 0.677 (45) & 0.644 (42) & 0.847 (33) & 0.890 (28) & \textbf{1.000} (0) \\
SAL(hybrid)       & 0.278 (86) & 0.455 (75) & 0.352 (86) & 0.515 (60) & 0.692 (53) & 0.730 (39) & 0.778 (35) & 0.809 (44) & 0.891 (25) & \textbf{1.000} (0) \\
LLM-Only          & \underline{0.788} (99) & 0.788 (99) & 0.788 (99) & 0.788 (99) & 0.788 (99) & 0.788 (99) & 0.788 (99) & 0.788 (99) & 0.788 (99) & 0.788 (99) \\
Random            & 0.788 (98) & \underline{0.789} (98) & \underline{0.789} (98) & \underline{0.789} (97) & \underline{0.792} (96) & 0.797 (94) & 0.806 (93) & 0.832 (89) & 0.898 (73) & \textbf{1.000} (0) \\
Naive             & \textit{0.789} (98) & \textit{0.790} (96) & \textit{0.791} (95) & \textit{0.794} (94) & \textit{0.802} (93) & \underline{0.818} (89) & \underline{0.847} (84) & 0.883 (72) & 0.912 (49) & \textbf{1.000} (0) \\
\textbf{LARA($n=1$)  }     & \textbf{0.834} (87) & \textbf{0.848} (79) & \textbf{0.860} (65) & \textbf{0.870} (55) & \textbf{0.875} (38) & \textbf{0.897} (19) & \textbf{0.917} (16) & \textbf{0.939} (13) & \textbf{0.970} (11) & \textbf{1.000} (0) \\
\textbf{LARA($n=3$)  }     & \textbf{0.843} (84) & \textbf{0.850} (72) & \textbf{0.864} (60) & \textbf{0.868} (50) & \textbf{0.874} (37) & \textbf{0.896} (18) & \textbf{0.921} (16) & \textit{0.939} (13) & \textit{0.970} (11) & \textbf{1.000} (0) \\
\textbf{LARA($n=N$) }    & \textbf{0.845} (82) & \textbf{0.856} (72) & \textbf{0.875} (60) & \textbf{0.882} (50) & \textbf{0.889} (36) & \textbf{0.904} (20) & \textbf{0.925} (18) & \textbf{0.950} (12) & \textbf{0.977} (9) & \textbf{1.000} (0) \\
\cline{1-11}
LLM-Only(rationale) & 0.843 (90) & 0.843 (90) & 0.843 (90) & 0.843 (90) & 0.843 (90) & 0.843 (90) & 0.843 (90) & 0.843 (90) & 0.843 (90) & 0.843 (90) \\
Random(rationale) & 0.843 (90) & 0.845 (90) & 0.845 (90) & 0.845 (89) & 0.849 (88) & 0.854 (88) & 0.860 (87) & 0.878 (84) & 0.922 (46) & \textbf{1.000} (1) \\
Naive(rationale) & 0.846 (89) & 0.846 (88) & 0.849 (88) & 0.850 (86) & 0.858 (86) & 0.876 (74) & 0.892 (41) & 0.935 (15) & 0.959 (11) & \textbf{1.000} (1) \\
\textbf{LARA(rationale)} & \textbf{0.847} (86) & \textbf{0.860} (79) & \textbf{0.871} (72) & \textbf{0.881} (60) & \textbf{0.899} (38) & \textbf{0.908} (19) & \textbf{0.938} (18) & \textbf{0.957} (13) & \textbf{0.986} (5) & \textbf{1.000} (1) \\
LLM-Only(utility) & 0.871 (75) & 0.871 (75) & 0.871 (75) & 0.871 (75) & 0.871 (75) & 0.871 (75) & 0.871 (75) & 0.871 (75) & 0.871 (75) & 0.871 (75) \\
Random(utility) & 0.871 (74) & 0.872 (74) & 0.874 (74) & 0.873 (73) & 0.879 (72) & 0.883 (71) & 0.890 (68) & 0.907 (56) & 0.927 (22) & \textbf{1.000} (1) \\
Naive(utility) & 0.871 (74) & 0.877 (74) & 0.878 (72) & 0.884 (68) & 0.896 (60) & 0.911 (40) & 0.933 (17) & 0.949 (12) & 0.968 (7) & \textbf{1.000} (1) \\
\textbf{LARA(utility)} & \textbf{0.877} (67) & \textbf{0.882} (50) & \textbf{0.886} (44) & \textbf{0.903} (23) & \textbf{0.919} (18) & \textbf{0.930} (19) & \textbf{0.937} (19) & \textbf{0.962} (13) & \textbf{0.988} (5) & \textbf{1.000} (1) \\
\cline{1-11}
LLM-Only (Llama-8B) & 0.676 (99) & 0.676 (99) & 0.676 (99) & 0.676 (99) & 0.676 (99) & 0.676 (99) & 0.676 (99) & 0.676 (99) & 0.676 (99) & 0.676 (99) \\
Random (Llama-8B) & 0.676 (99) & 0.676 (99) & 0.677 (99) & 0.680 (99) & 0.683 (99) & 0.687 (99) & 0.697 (99) & 0.717 (98) & 0.784 (96) & \textbf{1.000} (1) \\
Naive (Llama-8B) & 0.675 (99) & 0.676 (99) & 0.677 (99) & 0.680 (99) & 0.684 (99) & 0.691 (99) & 0.707 (98) & 0.723 (97) & 0.788 (95) & \textbf{1.000} (1) \\
\textbf{LARA (Llama-8B)} & \textbf{0.723} (91) & \textbf{0.754} (90) & \textbf{0.773} (84) & \textbf{0.786} (83) & \textbf{0.812} (77) & \textbf{0.845} (70) & \textbf{0.871} (66) & \textbf{0.890} (45) & \textbf{0.926} (18) & \textbf{1.000} (1) \\
\hline

\end{tabular}
\vspace{-15pt}
\label{tab:map_tau}
\end{table*}

\begin{table*}[t]
\caption{Kendall's Tau correlation scores (larger is better) for the ranked systems based on NDCG scores across different methods and different ratios of manual annotation budget. The first blocks show the main results, where for each column, the top \textit{three} highest scores are shown in bold, the fourth is in italics, and the fifth is underlined. For the methods based on the rationale and utility prompt, only the top score \textit{of each prompt} is shown in bold. For the methods based on Llama-8B, the top score is highlighted in bold. Values in parentheses show the Maximum Drop in the system rank (smaller is better).}
\centering
\small
\vspace{-10pt}
% We removed the "Dataset" column from the original specification, so change {ll|rrrrrrrrrr} to {l|rrrrrrrrrr}.
\begin{tabular}{l|rrrrrrrrrr}
\hline
% ==========================
% TREC Robust04 Section
% ==========================
\multicolumn{11}{c}{\textbf{TREC Robust04}} \\
\hline
& \multicolumn{10}{c}{\textbf{Ratio of Manual Annotation Budget ($B/|D|$)}} \\
\textbf{Method} & \textbf{$\frac{1}{512}$} & \textbf{$\frac{1}{256}$} & \textbf{$\frac{1}{128}$} & \textbf{$\frac{1}{64}$} & \textbf{$\frac{1}{32}$} & \textbf{$\frac{1}{16}$} & \textbf{$\frac{1}{8}$} & \textbf{$\frac{1}{4}$} & \textbf{$\frac{1}{2}$} & all \\
\hline
Depth-k & 0.662 (44) & 0.656 (40) & 0.699 (39) & 0.756 (37) & 0.793 (34) & 0.846 (26) & 0.898 (18) & 0.939 (12) & \underline{0.977} (5) & \textbf{1.000} (0) \\
MTF & 0.420 (60) & 0.442 (56) & 0.501 (59) & 0.535 (57) & 0.608 (49) & 0.611 (48) & 0.673 (41) & 0.727 (33) & 0.951 (10) & \textbf{1.000} (0) \\
MM-NS & 0.490 (57) & 0.490 (57) & 0.490 (57) & 0.577 (56) & 0.645 (44) & 0.622 (47) & 0.678 (41) & 0.730 (32) & 0.950 (10) & \textbf{1.000} (0) \\
CAL(human) & 0.662 (52) & 0.775 (31) & 0.779 (30) & 0.857 (26) & 0.852 (22) & 0.856 (34) & 0.914 (21) & 0.874 (21) & 0.914 (13) & \textbf{1.000} (0) \\
CAL(hybrid) & 0.695 (69) & 0.710 (54) & 0.787 (24) & 0.832 (25) & 0.834 (28) & 0.900 (23) & 0.868 (37) & 0.873 (28) & 0.919 (12) & \textbf{1.000} (0) \\
SAL(human) & 0.376 (54) & 0.665 (39) & 0.765 (43) & 0.799 (26) & 0.819 (39) & 0.860 (19) & 0.925 (17) & 0.892 (18) & 0.918 (11) & \textbf{1.000} (0) \\
SAL(hybrid) & 0.724 (45) & 0.663 (41) & 0.638 (55) & 0.770 (35) & 0.799 (34) & 0.820 (27) & 0.888 (17) & 0.914 (18) & 0.931 (11) & \textbf{1.000} (0) \\
LLM-Only & \underline{0.948} (22) & \underline{0.948} (22) & 0.948 (22) & \underline{0.948} (22) & \underline{0.948} (22) & \underline{0.948} (22) & \underline{0.948} (22) & \underline{0.948} (22) & 0.948 (22) & 0.948 (22) \\
Random & 0.948 (22) & 0.948 (22) & \underline{0.949} (22) & 0.948 (22) & 0.947 (22) & 0.946 (22) & 0.948 (22) & 0.948 (21) & 0.955 (21) & \textbf{1.000} (0) \\
Naive & \textit{0.948} (22) & \textbf{0.949} (22) & \textit{0.950} (22) & \textbf{0.951} (22) & \textit{0.950} (22) & \textit{0.959} (23) & \textit{0.965} (21) & \textit{0.971} (19) & \textit{0.993} (18) & \textbf{1.000} (0) \\
\textbf{LARA($n=1$)} & \textbf{0.949} (22) & \textit{0.949} (22) & \textbf{0.950} (22) & \textit{0.951} (22) & \textbf{0.953} (22) & \textbf{0.960} (22) & \textbf{0.968} (22) & \textbf{0.972} (21) & \textbf{0.994} (11) & \textbf{1.000} (0) \\
\textbf{LARA($n=3$)} & \textbf{0.949} (22) & \textbf{0.949} (22) & \textbf{0.950} (22) & \textbf{0.951} (22) & \textbf{0.953} (22) & \textbf{0.960} (22) & \textbf{0.968} (22) & \textbf{0.972} (21) & \textbf{0.994} (11) & \textbf{1.000} (0) \\
\textbf{LARA($n=N$)} & \textbf{0.949} (22) & \textbf{0.949} (22) & \textbf{0.950} (22) & \textbf{0.951} (22) & \textbf{0.957} (22) & \textbf{0.960} (22) & \textbf{0.968} (21) & \textbf{0.976} (18) & \textbf{0.993} (10) & \textbf{1.000} (0) \\
\cline{1-11}
LLM-Only(rationale) & 0.945 (23) & 0.945 (23) & 0.945 (23) & 0.945 (23) & 0.945 (23) & 0.945 (23) & 0.945 (23) & 0.945 (23) & 0.945 (23) & 0.945 (23) \\
Random(rationale) & 0.945 (23) & 0.945 (22) & 0.945 (22) & 0.946 (22) & 0.951 (22) & 0.954 (22) & 0.962 (21) & 0.970 (18) & 0.988 (10) & \textbf{1.000} (0) \\
Naive(rationale) & 0.945 (23) & 0.945 (23) & 0.945 (23) & 0.946 (22) & 0.952 (22) & 0.955 (22) & 0.962 (21) & 0.971 (18) & 0.987 (10) & \textbf{1.000} (0) \\
\textbf{LARA(rationale)} &  \textbf{0.945} (22) & \textbf{0.945} (22) & \textbf{0.946} (22) & \textbf{0.947} (22) & \textbf{0.953} (22) & \textbf{0.956} (22) & \textbf{0.964} (21) & \textbf{0.972} (18) & \textbf{0.989} (10) & \textbf{1.000} (0) \\
LLM-Only(utility) & 0.947 (21) & 0.947 (21) & 0.947 (21) & 0.947 (21) & 0.947 (21) & 0.947 (21) & 0.947 (21) & 0.947 (21) & 0.947 (21) & 0.947 (21) \\
Random(utility) & 0.946 (21) & 0.945 (21) & 0.947 (20) & 0.948 (20) & 0.953 (20) & 0.957 (20) & 0.964 (19) & 0.978 (17) & 0.993 (10) & \textbf{1.000} (0) \\
Naive(utility) & 0.947 (21) & 0.947 (21) & 0.946 (20) & 0.947 (20) & 0.954 (20) & 0.956 (20) & 0.963 (19) & 0.979 (17) & 0.994 (10) & \textbf{1.000} (0) \\
\textbf{LARA(utility)} & \textbf{0.947} (21) & \textbf{0.947} (21) & \textbf{0.948} (20) & \textbf{0.949} (20) & \textbf{0.955} (20) & \textbf{0.958} (20) & \textbf{0.966} (19) & \textbf{0.980} (17) & \textbf{0.995} (10) & \textbf{1.000} (0) \\
\cline{1-11}
LLM-Only (Llama-8B) & 0.894 (33) & 0.894 (33) & 0.894 (33) & 0.894 (33) & 0.894 (33) & 0.894 (33) & 0.894 (33) & 0.894 (33) & 0.894 (33) & 0.894 (33) \\
Random (Llama-8B) & 0.894 (33) & 0.894 (33) & 0.894 (33) & 0.894 (33) & 0.894 (33) & 0.894 (33) & 0.896 (32) & 0.898 (32) & 0.905 (28) & \textbf{1.000} (0) \\
Naive (Llama-8B) & 0.894 (33) & 0.894 (33) & 0.894 (33) & 0.897 (33) & 0.899 (33) & 0.901 (32) & 0.905 (32) & 0.906 (29) & 0.919 (24) & \textbf{1.000} (0) \\
\textbf{LARA (Llama-8B)} & \textbf{0.896} (33) & \textbf{0.896} (33) & \textbf{0.896} (33) & \textbf{0.900} (33) & \textbf{0.900} (33) & \textbf{0.904} (32) & \textbf{0.908} (31) & \textbf{0.911} (27) & \textbf{0.924} (24) & \textbf{1.000} (0) \\
\hline
% ==========================
% TREC COVID Section
% ==========================
\multicolumn{11}{c}{\textbf{TREC COVID}} \\
\hline
& \multicolumn{10}{c}{\textbf{Ratio of Manual Annotation Budget ($B/|D|$)}} \\
\textbf{Method} & \textbf{$\frac{1}{512}$} & \textbf{$\frac{1}{256}$} & \textbf{$\frac{1}{128}$} & \textbf{$\frac{1}{64}$} & \textbf{$\frac{1}{32}$} & \textbf{$\frac{1}{16}$} & \textbf{$\frac{1}{8}$} & \textbf{$\frac{1}{4}$} & \textbf{$\frac{1}{2}$} & all \\
\hline
Depth-k & 0.636 (62) & 0.673 (66) & 0.704 (54) & 0.771 (41) & 0.818 (29) & 0.860 (26) & 0.911 (16) & 0.953 (14) & 0.963 (12) & \textbf{1.000} (0) \\
MTF & 0.651 (65) & 0.614 (40) & 0.681 (30) & 0.749 (35) & 0.833 (28) & 0.870 (25) & 0.908 (25) & 0.949 (16) & 0.963 (12) & \textbf{1.000} (0) \\
MM-NS & 0.572 (81) & 0.355 (68) & 0.734 (41) & 0.806 (33) & 0.853 (28) & 0.878 (25) & 0.873 (25) & 0.946 (16) & 0.963 (12) & \textbf{1.000} (0) \\
CAL(human) & 0.452 (80) & 0.655 (27) & 0.657 (31) & 0.860 (28) & 0.875 (42) & 0.921 (11) & 0.937 (13) & 0.954 (8) & \textit{0.981} (4) & \textbf{1.000} (0) \\
CAL(hybrid) & 0.751 (43) & 0.693 (71) & 0.794 (44) & 0.663 (35) & 0.760 (42) & 0.824 (23) & 0.901 (29) & 0.898 (8) & \underline{0.978} (4) & \textbf{1.000} (0) \\
SAL(human) & 0.456 (56) & 0.541 (54) & 0.664 (79) & 0.348 (74) & 0.498 (59) & 0.781 (30) & 0.817 (20) & 0.870 (16) & 0.947 (10) & \textbf{1.000} (0) \\
SAL(hybrid) & 0.285 (82) & 0.373 (73) & 0.406 (50) & 0.742 (47) & 0.555 (22) & 0.765 (24) & 0.806 (19) & 0.882 (17) & 0.972 (7) & \textbf{1.000} (0) \\
LLM-Only & \textit{0.950} (19) & \underline{0.946} (19) & \textit{0.950} (19) & \underline{0.950} (19) & \textit{0.950} (19) & 0.950 (19) & 0.950 (19) & 0.950 (19) & 0.950 (19) & 0.950 (19) \\
Random & \underline{0.949} (19) & \textit{0.946} (19) & \underline{0.950} (19) & \textit{0.951} (18) & \underline{0.949} (19) & \underline{0.952} (18) & \underline{0.954} (15) & \underline{0.959} (15) & 0.977 (12) & \textbf{1.000} (0) \\
Naive & 0.942 (19) & 0.942 (19) & 0.944 (19) & 0.947 (18) & 0.949 (18) & \textbf{0.957} (18) & \textbf{0.961} (16) & \textit{0.965} (16) & 0.972 (15) & \textbf{1.000} (0) \\
\textbf{LARA($n=1$)} & \textbf{0.951} (19) & \textbf{0.950} (18) & \textbf{0.951} (18) & \textbf{0.952} (18) & \textbf{0.955} (18) & \textbf{0.956} (17) & \textit{0.954} (15) & \textbf{0.970} (15) & \textbf{0.986} (11) & \textbf{1.000} (0) \\
\textbf{LARA($n=3$)} & \textbf{0.951} (19) & \textbf{0.950} (18) & \textbf{0.951} (18) & \textbf{0.952} (18) & \textbf{0.955} (18) & \textbf{0.956} (17) & \textbf{0.954} (15) & \textbf{0.970} (15) & \textbf{0.986} (11) & \textbf{1.000} (0) \\
\textbf{LARA($n=N$)} & \textbf{0.950} (19) & \textbf{0.951} (18) & \textbf{0.952} (18) & \textbf{0.953} (18) & \textbf{0.955} (20) & \textit{0.955} (15) & \textbf{0.965} (15) & \textbf{0.977} (15) & \textbf{0.990} (11) & \textbf{1.000} (0) \\
\cline{1-11}
LLM-Only(rationale) & 0.913 (22) & 0.913 (22) & 0.913 (22) & 0.913 (22) & 0.913 (22) & 0.913 (22) & 0.913 (22) & 0.913 (22) & 0.913 (22) & 0.913 (22) \\
Random(rationale) & 0.913 (22) & 0.913 (22) & 0.914 (22) & 0.915 (21) & 0.914 (22) & 0.916 (21) & 0.922 (20) & 0.935 (15) & 0.955 (14) & \textbf{1.000} (0) \\
Naive(rationale) & 0.913 (22) & 0.914 (22) & 0.916 (21) & 0.917 (21) & 0.919 (21) & 0.924 (20) & 0.929 (20) & 0.937 (16) & 0.954 (15) & \textbf{1.000} (0) \\
\textbf{LARA(rationale)} & \textbf{0.915} (22) & \textbf{0.915} (22) & \textbf{0.917} (22) & \textbf{0.918} (21) & \textbf{0.921} (21) & \textbf{0.925} (21) & \textbf{0.930} (18) & \textbf{0.943} (15) & \textbf{0.961} (15) & \textbf{1.000} (0) \\
LLM-Only(utility) & 0.929 (15) & 0.929 (15) & 0.929 (15) & 0.929 (15) & 0.929 (15) & 0.929 (15) & 0.929 (15) & 0.929 (15) & 0.929 (15) & 0.929 (15) \\
Random(utility) & 0.928 (15) & 0.930 (15) & 0.929 (16) & 0.931 (15) & 0.930 (15) & 0.933 (15) & 0.937 (15) & 0.950 (15) & 0.967 (12) & \textbf{1.000} (0) \\
Naive(utility) & 0.929 (15) & 0.930 (15) & 0.929 (16) & 0.930 (16) & 0.931 (15) & 0.935 (15) & 0.937 (15) & 0.951 (15) & 0.971 (12) & \textbf{1.000} (0) \\
\textbf{LARA(utility)} & \textbf{0.930} (15) & \textbf{0.931} (15) & \textbf{0.932} (15) & \textbf{0.932} (15) & \textbf{0.936} (15) & \textbf{0.940} (15) & \textbf{0.947} (15) & \textbf{0.957} (15) & \textbf{0.973} (11) & \textbf{1.000} (0) \\
\cline{1-11}
LLM-Only (Llama-8B) & 0.869 (31) & 0.869 (31) & 0.869 (31) & 0.869 (31) & 0.869 (31) & 0.869 (31) & 0.869 (31) & 0.869 (31) & 0.869 (31) & 0.869 (31) \\
Random (Llama-8B) & 0.869 (31) & 0.871 (31) & 0.870 (31) & 0.872 (31) & 0.872 (31) & 0.876 (31) & 0.883 (31) & 0.897 (28) & 0.922 (20) & \textbf{1.000} (0) \\
Naive (Llama-8B) & 0.870 (31) & 0.870 (31) & 0.871 (31) & 0.874 (31) & \textbf{0.876} (31) & 0.881 (31) & 0.890 (31) & 0.905 (29) & 0.946 (16) & \textbf{1.000} (0) \\
\textbf{LARA (Llama-8B)} & \textbf{0.870} (31) & \textbf{0.871} (31) & \textbf{0.872} (31) & \textbf{0.875} (31) & 0.876 (31) & \textbf{0.883} (31) & \textbf{0.895} (30) & \textbf{0.917} (27) & \textbf{0.956} (15) & \textbf{1.000} (0) \\
\hline
\end{tabular}
\vspace{-15pt}
\label{tab:ndcg_tau}
\end{table*}

\begin{figure*}[tp]
    \vspace{-6pt}
    \begin{tabular}{cc}
        \includegraphics[width=3.00in, keepaspectratio]{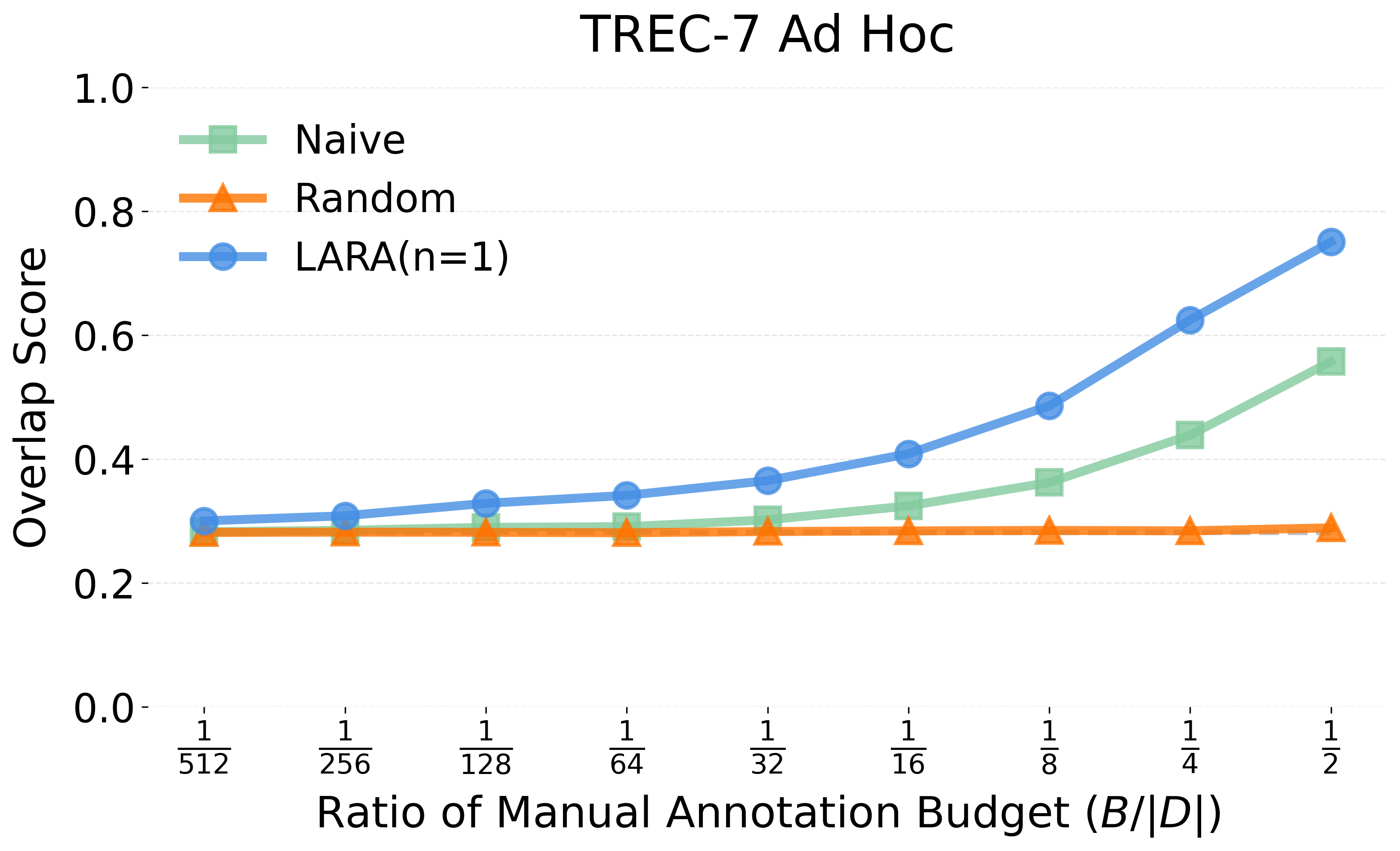} &
        \includegraphics[width=3.00in, keepaspectratio]{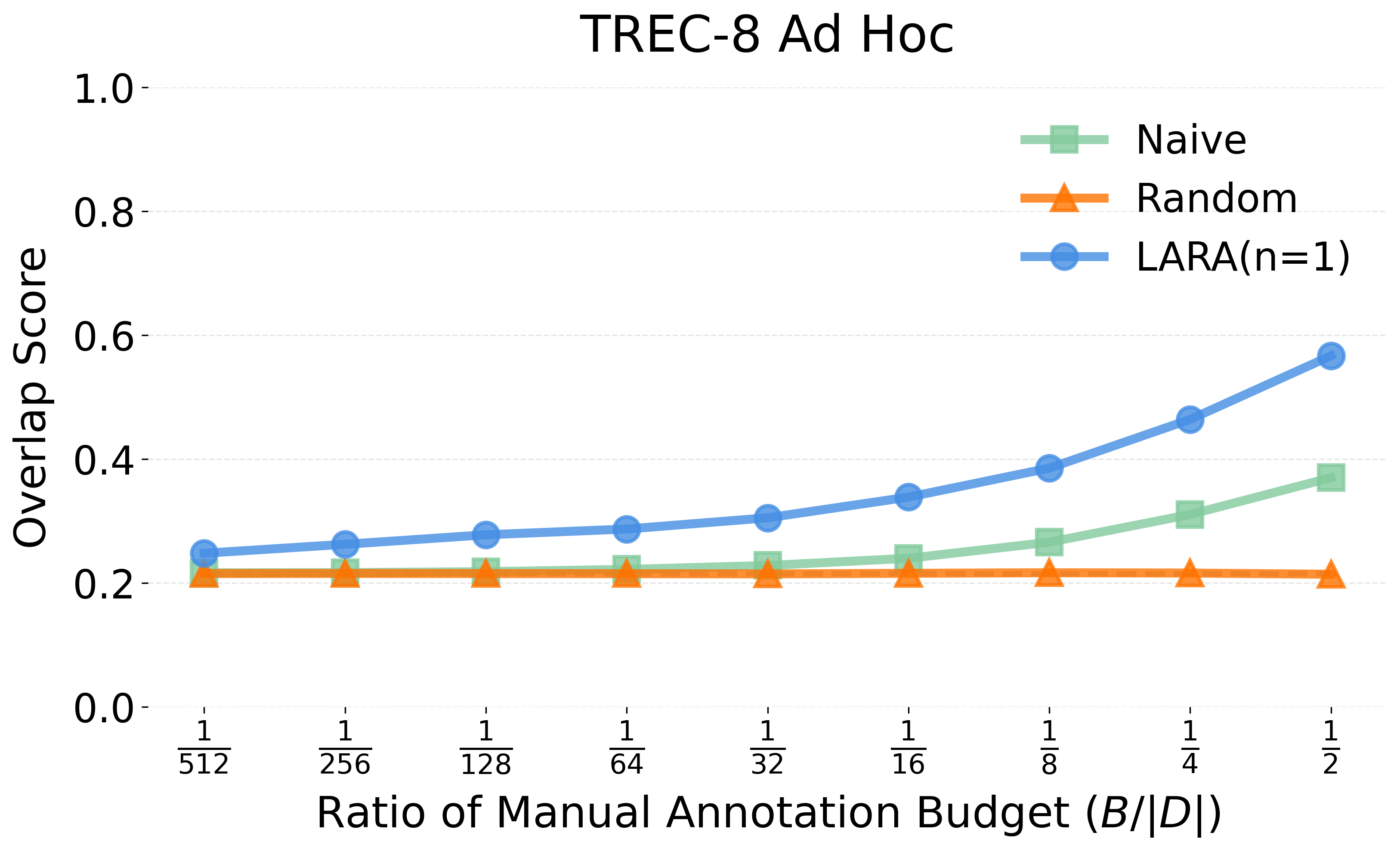} \\ [3pt]
        \includegraphics[width=3.00in, keepaspectratio]{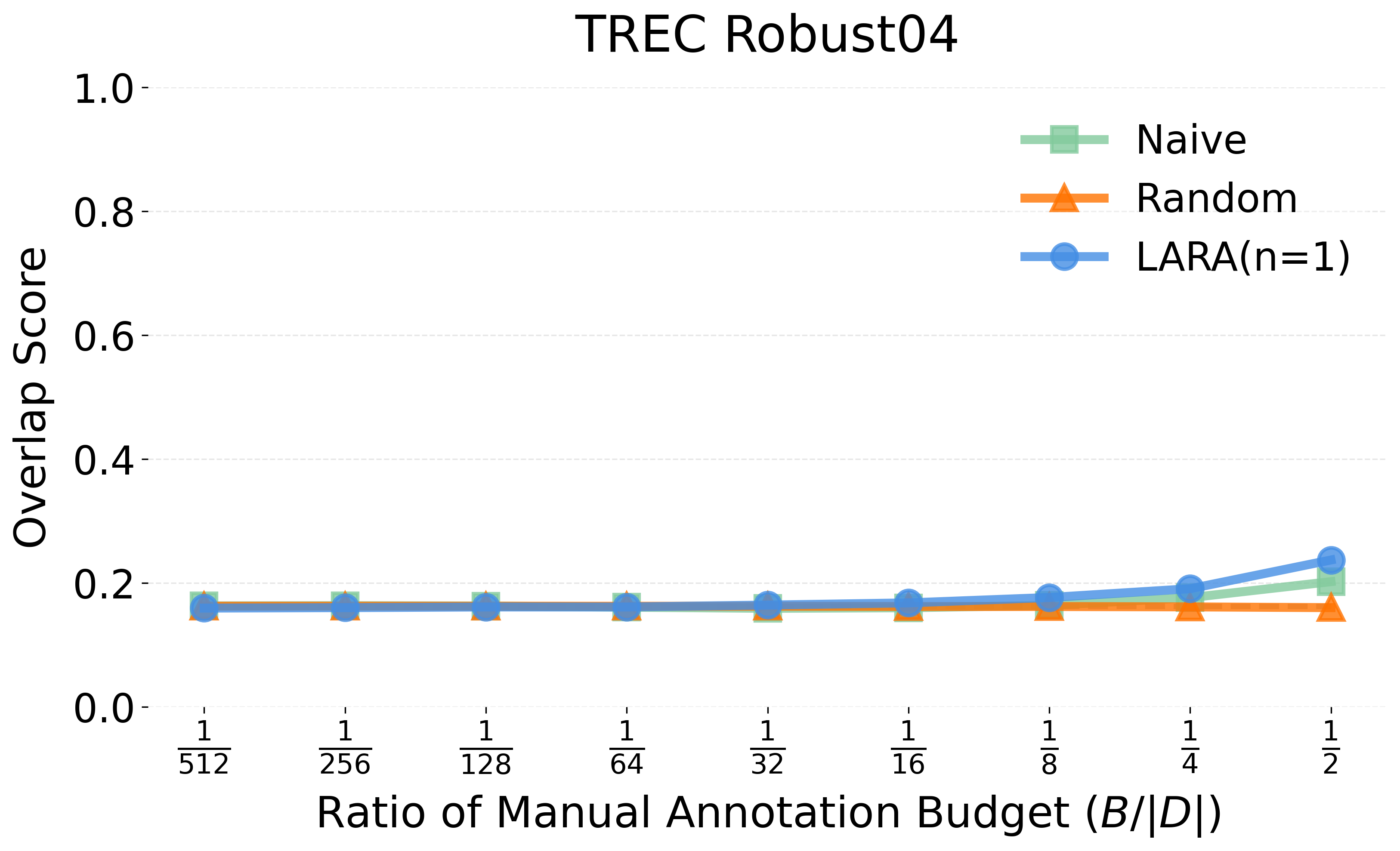} &
        \includegraphics[width=3.00in, keepaspectratio]{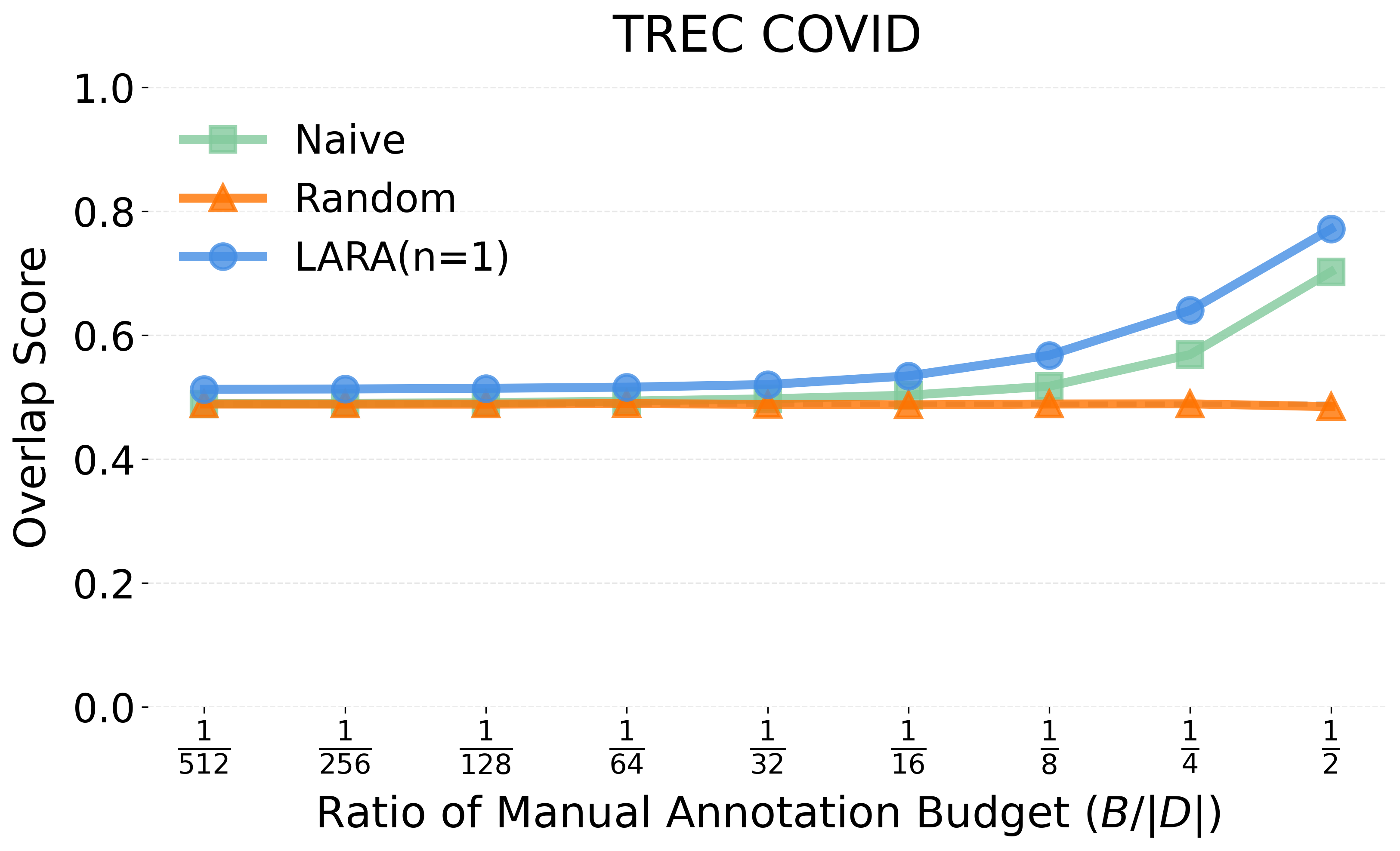}
    \end{tabular}
    \vspace{-10pt}
    \Description{Four line charts showing overlap score versus manual‐annotation
    budget ratio for three methods (Naive, Random, LARA) on TREC-7, TREC-8,
    Robust04 and TREC-COVID. In every dataset the blue LARA curve stays
    consistently above the green Naive and orange Random curves, and the
    performance gap widens as budget increases, indicating LARA yields the
    fewest LLM-annotation errors.}
    \caption{Comparison of overlap scores of LLM-based methods on different budgets on TREC-7 Ad Hoc (top left), TREC-8 Ad Hoc (top right), TREC Robust04 (bottom left), and TREC-COVID (bottom right).}
    \label{fig:enter-label}
    \vspace{-10pt}
\end{figure*}

\section{Experimental Results}
\subsection{Main Results}
The main results of the binary relevance experiments on the TREC-7 and TREC-8 Ad Hoc datasets are summarized in Table~\ref{tab:map_tau}, and the main results of the graded relevance experiments on the TREC Robust04 and TREC COVID datasets are summarized in Table~\ref{tab:ndcg_tau}. We separate the results to distinguish the binary relevance experiments evaluated using the MAP scores clearly from the graded relevance experiments evaluated using the NDCG scores.

From Tables~\ref{tab:map_tau} and \ref{tab:ndcg_tau}, we observe that our LARA method outperforms all alternatives on almost any budget constraints on all datasets. We analyze more by answering the following questions.\\

\noindent
\textbf{\textit{How does LARA perform compared to existing methods based on manual annotations?}}
Compared to the baseline methods that use only manual judgments (i.e., Depth-\textit{k} Pooling, MTF, MM-NS, CAL(human), and SAL(human)), LARA almost always performs better, particularly when the budget is low. This result was anticipated because while human annotations are always the ground truth, human-based methods lack the annotation data to make the evaluation stable on a low budget. The baseline hybrid methods (i.e., CAL(hybrid) and SAL(hybrid)) have similar results as those that use only human labels since the regression model can be trained only on small data when the budget is low. Although the Depth-$k$ pooling method does outperform some LARA methods in a few of the scenarios in the binary relevance experiments, we observe that its performance is not consistent, particularly on lower budgets. Moreover, the difference in performance between LARA and the manual-based methods is expected to be larger in real-life applications, since LLMs would be able to assess much more than the limit in these test collections. For example, TREC-8 Ad Hoc is a test collection with depth-100 pooling, but if we use LLMs, we could collect documents from depth-1000 or even from the entire document collection. Indeed, we observe that the LLM-based methods dominate the manual methods on the large TREC Robust04 test collection in Table~\ref{tab:ndcg_tau}, especially on a lower budget.\\

\noindent
\textbf{\textit{How does LARA perform compared to the case of only using LLM?}}
Compared to the case with only LLM annotations (i.e., LLM-Only), LARA's performance is only marginally better at first, but the difference widens as the budget increases. This is because LARA chooses effective data points to give ground truth judgments, which decreases the number of annotation errors.\\

\noindent
\textbf{\textit{How does LARA perform compared to other methods that use both LLM and manual assessments?}}
The Naive and Random methods, which use both LLM and human assessments, almost constantly perform better than the other baseline methods, particularly on a low budget. This alone shows the merit of using both LLM and human judgments. Nonetheless, our LARA method outperforms the Naive and Random methods in most settings, which confirms that our method of calibrating the predictions enhances the use of LLMs for building test collections even more. We elaborate further on this in the exploratory experiments in Section~\ref{sec:exploratory}.\\

\noindent
\textbf{\textit{How does the number of annotators affect the performance of LARA?}}
The LARA method with different numbers of annotators (i.e., LARA($n=1$), LARA($n=3$), and LARA($n=N$)), perhaps surprisingly, have almost identical performance on all datasets and budgets, with LARA($n=N$) slightly outperforming the other two in many scenarios. We conjecture that this occurs because as the number of annotators $n$ increases, the number of ground truth annotations would be distributed fairly among each group of topics. This allows each topic to have similar qualities of evaluation results, which helped increase the overall performance. In contrast, when the number of annotators $n$ is small, although the algorithm would be able to choose the most likely-uncertain data point from a wider range of topics, the evaluation performance could vary greatly from topic to topic, since some topics could have very accurate evaluations while others could have poor performance due to differences in manual annotation distribution. This interesting result supports the practicality of our LARA method. We would thus recommend using LARA($n=N$) since it can handle any number of assessors.\\

\noindent
\textbf{\textit{How does LARA perform in terms of Maximum Drop?}}
In the tables of main results, we also report the Maximum Drop values in the parentheses. We observe that LARA constantly has one of the lowest drops for all budgets in all datasets. Specifically, compared to the manual methods, LARA's Maximum Drop is much lower under lower budgets and performs mostly better also under higher budgets. Importantly, only using LLM judgments may lead to very high drops although achieving fairly high Kendall's $\tau$ scores; this trend is particularly visible in the TREC-8 Ad Hoc experiment with a Maximum Drop of 99. Thus, we argue that it can be risky to rely only on manual judgments on low budgets or to rely only on LLM judgments. We observe that LARA is almost always safer than any alternatives on all budget constraints.\\

\noindent
\textbf{\textit{How does the prompt affect the quality of LARA?}}
As an ablation study, we also show results with the rationale prompt~\citep{upadhyay2024largescalestudyrelevanceassessments} and the utility prompt~\citep{thomas2024llmreljudge}. From the results, we observe that LARA outperforms the alternative LLM-based methods on the same prompt and the manual baselines on almost all tested settings. This implies the robustness and consistency of our LARA algorithm.\\

\noindent
\textbf{\textit{How does the performance of the LLM affect the quality of LARA?}}
We show the results of LLM-based methods using Llama-3.1-8B-Instruct, a lightweight variant of Llama-3.1-70B-Instruct. Comparing the LLM-Only methods, we first know that the 8B model performs constantly worse than the 70B model. Nonetheless, while LARA (Llama-8B) generally never outperforms LARA with the 70B model, LARA (Llama-8B) outperforms the rest of the 8B-based methods, again showing the effectiveness of LARA. Compared to the existing manual methods, LARA (Llama-8B) performs better under low-budget constraints for all datasets. Thus, when building a test collection under a low budget, LARA is a good option even with a small LLM, but we recommend using a better LLM if available.

\subsection{Exploratory Results: LLM Annotation Error}
\label{sec:exploratory}
One crucial merit of our algorithm, other than selecting effective documents to manually annotate, is that it can also minimize LLM annotation errors. To visualize this, we compare LARA($n=1$) with Naive and Random methods and show the difference in annotation errors. We do not show LARA($n=3$) and LARA($n=N$) here, because the results were almost identical to LARA($n=1$), so the lines were completely on top of each other. We show the overlap scores of only the LLM-annotated data points. The manually-annotated data points were not considered here, since these data points are always correct and thus not informative.

From the results in Figure~\ref{fig:enter-label}, we can observe that LARA consistently has a better overlap score than the alternatives on all budgets, which shows that LARA is able to minimize the LLM annotation error by calibrating the LLM judgments with the learned model and by giving truth labels to the data points with the highest expected error. The Naive method also improves its performance because this method also tries to manually annotate the most uncertain data points. Nonetheless, LARA constantly outperforms it in terms of LLM annotation accuracy since the predictions are based on the calibration model $\hat{p}(y|\pi)$ instead of naively using $\pi$.

\section{Discussions and Future Directions}
\label{sec: discussions}
As a potentially effective alternative to LARA, one might consider actively fine-tuning the LLM after each batch of human annotations to potentially improve its predictions. However, this approach poses practical challenges. Repeatedly fine-tuning an LLM each time a batch of annotations is collected would be time-consuming, requiring assessors to wait for the model to update and to re-annotate all the data points before proceeding to the next batch\footnote{LLM annotation of TREC Robust04 took about 5-6 days; others took about 2 days.}. This would greatly disrupt the annotation workflow and reduce efficiency. Instead, our current approach only requires the LLM to output \textit{once} and focuses on a calibration model that can be updated rapidly\footnote{In the experiments, LARA took only about 1-10 seconds to build a full test collection on any budget constraints. Note, however, that we also need to consider the time that takes to assess manually in practice.}.

Nonetheless, it may be beneficial to fine-tune the LLM \textit{pre-deployment}, using historical or domain-specific data~\citep{meng2024queryperformancepredictionusing}. Such pre-deployment tuning could yield a more accurate starting point for LARA, as we see from the results that the quality of LLMs affects the performance of LARA. Exploring the trade-offs between continuous retraining, pre-deployment fine-tuning, and on-the-fly calibration is an interesting avenue for future research. As future work, it would also be valuable to investigate a LARA-like hybrid method on other annotation tasks like nugget matching~\citep{Takehi2023nuggetlevel, nenkova2004pyramid}, which aims to annotate whether some information nugget can be obtained from a text, and e-Discovery~\citep{doug2018ediscovery, cormack2104tar}, which aims to accurately collect relevant documents for use in legal cases.

\section{Conclusions}
This paper investigates an effective and efficient way of building a test collection, particularly using both LLM and limited manual assessments. Relying only on limited manual assessments leads to unstable evaluations due to limited collection size, and relying only on LLM judgments is risky and unreliable. Thus, we propose \textbf{L}LM-\textbf{A}ssisted \textbf{R}elevance \textbf{A}ssessments (\textbf{LARA}) to build a risk-minimized yet large test collection. LARA actively learns to calibrate LLM relevance predictions by strategically selecting documents for human annotation. This learned calibration model guides which documents should be manually reviewed and corrects the LLM predictions for the remaining documents. Experiments on various datasets demonstrate that LARA consistently outperforms all the tested alternatives under almost any budget constraint, suggesting that LARA could help venues like TREC, CLEF, and NTCIR obtain large, stable, and reliable test collections even when the budget for manual annotations is limited.

\bibliographystyle{ACM-Reference-Format}
\balance
\bibliography{ref}


\begin{thebibliography}{43}

%%% the Web address.

\ifx \showCODEN    \undefined \def \showCODEN     #1{\unskip}     \fi
\ifx \showDOI      \undefined \def \showDOI       #1{#1}\fi
\ifx \showISBNx    \undefined \def \showISBNx     #1{\unskip}     \fi
\ifx \showISBNxiii \undefined \def \showISBNxiii  #1{\unskip}     \fi
\ifx \showISSN     \undefined \def \showISSN      #1{\unskip}     \fi
\ifx \showLCCN     \undefined \def \showLCCN      #1{\unskip}     \fi
\ifx \shownote     \undefined \def \shownote      #1{#1}          \fi
\ifx \showarticletitle \undefined \def \showarticletitle #1{#1}   \fi
\ifx \showURL      \undefined \def \showURL       {\relax}        \fi
% invisible to TeX
\providecommand\bibfield[2]{#2}
\providecommand\bibinfo[2]{#2}
\providecommand\natexlab[1]{#1}
\providecommand\showeprint[2][]{arXiv:#2}

\bibitem[Abbasiantaeb et~al\mbox{.}(2024)]%
        {abbasiantaeb2024uselargelanguagemodels}
\bibfield{author}{\bibinfo{person}{Zahra Abbasiantaeb}, \bibinfo{person}{Chuan Meng}, \bibinfo{person}{Leif Azzopardi}, {and} \bibinfo{person}{Mohammad Aliannejadi}.} \bibinfo{year}{2024}\natexlab{}.
\newblock \bibinfo{title}{Can We Use Large Language Models to Fill Relevance Judgment Holes?}
\newblock
\newblock
\showeprint[arxiv]{2405.05600}~[cs.IR]
\urldef\tempurl%
\url{https://arxiv.org/abs/2405.05600}
\showURL{%
\tempurl}


\bibitem[Alaofi et~al\mbox{.}(2024)]%
        {alaofi2024fooledllm}
\bibfield{author}{\bibinfo{person}{Marwah Alaofi}, \bibinfo{person}{Paul Thomas}, \bibinfo{person}{Falk Scholer}, {and} \bibinfo{person}{Mark Sanderson}.} \bibinfo{year}{2024}\natexlab{}.
\newblock \showarticletitle{LLMs can be Fooled into Labelling a Document as Relevant: best cafe near me; this paper is perfectly relevant}. In \bibinfo{booktitle}{\emph{Proceedings of the 2024 Annual International ACM SIGIR Conference on Research and Development in Information Retrieval in the Asia Pacific Region}} (Tokyo, Japan) \emph{(\bibinfo{series}{SIGIR-AP 2024})}. \bibinfo{publisher}{Association for Computing Machinery}, \bibinfo{address}{New York, NY, USA}, \bibinfo{pages}{32–41}.
\newblock
\showISBNx{9798400707247}
\urldef\tempurl%
\url{https://doi.org/10.1145/3673791.3698431}
\showDOI{\tempurl}


\bibitem[Alonso and Baeza-Yates(2011)]%
        {alonso2011crowd}
\bibfield{author}{\bibinfo{person}{Omar Alonso} {and} \bibinfo{person}{Ricardo Baeza-Yates}.} \bibinfo{year}{2011}\natexlab{}.
\newblock \showarticletitle{Design and implementation of relevance assessments using crowdsourcing}. In \bibinfo{booktitle}{\emph{Proceedings of the 33rd European Conference on Advances in Information Retrieval}} (Dublin, Ireland) \emph{(\bibinfo{series}{ECIR'11})}. \bibinfo{publisher}{Springer-Verlag}, \bibinfo{address}{Berlin, Heidelberg}, \bibinfo{pages}{153–164}.
\newblock
\showISBNx{9783642201608}


\bibitem[Aslam et~al\mbox{.}(2006)]%
        {aslam2006statsmethod}
\bibfield{author}{\bibinfo{person}{Javed~A. Aslam}, \bibinfo{person}{Virgil Pavlu}, {and} \bibinfo{person}{Emine Yilmaz}.} \bibinfo{year}{2006}\natexlab{}.
\newblock \showarticletitle{A statistical method for system evaluation using incomplete judgments}. In \bibinfo{booktitle}{\emph{Proceedings of the 29th Annual International ACM SIGIR Conference on Research and Development in Information Retrieval}} (Seattle, Washington, USA) \emph{(\bibinfo{series}{SIGIR '06})}. \bibinfo{publisher}{Association for Computing Machinery}, \bibinfo{address}{New York, NY, USA}, \bibinfo{pages}{541–548}.
\newblock
\showISBNx{1595933697}
\urldef\tempurl%
\url{https://doi.org/10.1145/1148170.1148263}
\showDOI{\tempurl}


\bibitem[Aslam and Yilmaz(2007)]%
        {aslam2007inferring}
\bibfield{author}{\bibinfo{person}{Javed~A. Aslam} {and} \bibinfo{person}{Emine Yilmaz}.} \bibinfo{year}{2007}\natexlab{}.
\newblock \showarticletitle{Inferring document relevance from incomplete information}. In \bibinfo{booktitle}{\emph{Proceedings of the Sixteenth ACM Conference on Conference on Information and Knowledge Management}} (Lisbon, Portugal) \emph{(\bibinfo{series}{CIKM '07})}. \bibinfo{publisher}{Association for Computing Machinery}, \bibinfo{address}{New York, NY, USA}, \bibinfo{pages}{633–642}.
\newblock
\showISBNx{9781595938039}
\urldef\tempurl%
\url{https://doi.org/10.1145/1321440.1321529}
\showDOI{\tempurl}


\bibitem[Bailey et~al\mbox{.}(2008)]%
        {bailey2008gold}
\bibfield{author}{\bibinfo{person}{Peter Bailey}, \bibinfo{person}{Nick Craswell}, \bibinfo{person}{Ian Soboroff}, \bibinfo{person}{Paul Thomas}, \bibinfo{person}{Arjen~P. de Vries}, {and} \bibinfo{person}{Emine Yilmaz}.} \bibinfo{year}{2008}\natexlab{}.
\newblock \showarticletitle{Relevance assessment: are judges exchangeable and does it matter}. In \bibinfo{booktitle}{\emph{Proceedings of the 31st Annual International ACM SIGIR Conference on Research and Development in Information Retrieval}} (Singapore, Singapore) \emph{(\bibinfo{series}{SIGIR '08})}. \bibinfo{publisher}{Association for Computing Machinery}, \bibinfo{address}{New York, NY, USA}, \bibinfo{pages}{667–674}.
\newblock
\showISBNx{9781605581644}
\urldef\tempurl%
\url{https://doi.org/10.1145/1390334.1390447}
\showDOI{\tempurl}


\bibitem[Clarke and Dietz(2024)]%
        {clarke2024llmbasedrelevanceassessmentcant}
\bibfield{author}{\bibinfo{person}{Charles L.~A. Clarke} {and} \bibinfo{person}{Laura Dietz}.} \bibinfo{year}{2024}\natexlab{}.
\newblock \bibinfo{title}{LLM-based relevance assessment still can't replace human relevance assessment}.
\newblock
\newblock
\showeprint[arxiv]{2412.17156}~[cs.IR]
\urldef\tempurl%
\url{https://arxiv.org/abs/2412.17156}
\showURL{%
\tempurl}


\bibitem[Clough et~al\mbox{.}(2013)]%
        {clough2013crowdsource}
\bibfield{author}{\bibinfo{person}{Paul Clough}, \bibinfo{person}{Mark Sanderson}, \bibinfo{person}{Jiayu Tang}, \bibinfo{person}{Tim Gollins}, {and} \bibinfo{person}{Amy Warner}.} \bibinfo{year}{2013}\natexlab{}.
\newblock \showarticletitle{Examining the Limits of Crowdsourcing for Relevance Assessment}.
\newblock \bibinfo{journal}{\emph{IEEE Internet Computing}} \bibinfo{volume}{17}, \bibinfo{number}{4} (\bibinfo{year}{2013}), \bibinfo{pages}{32--38}.
\newblock
\urldef\tempurl%
\url{https://doi.org/10.1109/MIC.2012.95}
\showDOI{\tempurl}


\bibitem[Cormack and Grossman(2014)]%
        {cormack2104tar}
\bibfield{author}{\bibinfo{person}{Gordon~V. Cormack} {and} \bibinfo{person}{Maura~R. Grossman}.} \bibinfo{year}{2014}\natexlab{}.
\newblock \showarticletitle{Evaluation of machine-learning protocols for technology-assisted review in electronic discovery}. In \bibinfo{booktitle}{\emph{Proceedings of the 37th International ACM SIGIR Conference on Research \& Development in Information Retrieval}} (Gold Coast, Queensland, Australia) \emph{(\bibinfo{series}{SIGIR '14})}. \bibinfo{publisher}{Association for Computing Machinery}, \bibinfo{address}{New York, NY, USA}, \bibinfo{pages}{153–162}.
\newblock
\showISBNx{9781450322577}
\urldef\tempurl%
\url{https://doi.org/10.1145/2600428.2609601}
\showDOI{\tempurl}


\bibitem[Cormack and Grossman(2016)]%
        {cormack2016scal}
\bibfield{author}{\bibinfo{person}{Gordon~V. Cormack} {and} \bibinfo{person}{Maura~R. Grossman}.} \bibinfo{year}{2016}\natexlab{}.
\newblock \showarticletitle{Scalability of Continuous Active Learning for Reliable High-Recall Text Classification}. In \bibinfo{booktitle}{\emph{Proceedings of the 25th ACM International on Conference on Information and Knowledge Management}} (Indianapolis, Indiana, USA) \emph{(\bibinfo{series}{CIKM '16})}. \bibinfo{publisher}{Association for Computing Machinery}, \bibinfo{address}{New York, NY, USA}, \bibinfo{pages}{1039–1048}.
\newblock
\showISBNx{9781450340731}
\urldef\tempurl%
\url{https://doi.org/10.1145/2983323.2983776}
\showDOI{\tempurl}


\bibitem[Cormack and Mojdeh(2009)]%
        {cormack2009cal}
\bibfield{author}{\bibinfo{person}{G.~V. Cormack} {and} \bibinfo{person}{M. Mojdeh}.} \bibinfo{year}{2009}\natexlab{}.
\newblock \showarticletitle{Machine learning for information retrieval: {TREC} 2009 {Web}, {Relevance Feedback} and {Legal Tracks}}. In \bibinfo{booktitle}{\emph{The Eighteenth Text REtrieval Conference ({TREC} 2009)}}.
\newblock


\bibitem[Cormack et~al\mbox{.}(1998)]%
        {cormack1998mtf}
\bibfield{author}{\bibinfo{person}{Gordon~V. Cormack}, \bibinfo{person}{Christopher~R. Palmer}, {and} \bibinfo{person}{Charles L.~A. Clarke}.} \bibinfo{year}{1998}\natexlab{}.
\newblock \showarticletitle{Efficient construction of large test collections}. In \bibinfo{booktitle}{\emph{Proceedings of the 21st Annual International ACM SIGIR Conference on Research and Development in Information Retrieval}} (Melbourne, Australia) \emph{(\bibinfo{series}{SIGIR '98})}. \bibinfo{publisher}{Association for Computing Machinery}, \bibinfo{address}{New York, NY, USA}, \bibinfo{pages}{282–289}.
\newblock
\showISBNx{1581130155}
\urldef\tempurl%
\url{https://doi.org/10.1145/290941.291009}
\showDOI{\tempurl}


\bibitem[Craswell et~al\mbox{.}(2024)]%
        {craswell2024deeplearning}
\bibfield{author}{\bibinfo{person}{Nick Craswell}, \bibinfo{person}{Bhaskar Mitra}, \bibinfo{person}{Emine Yilmaz}, \bibinfo{person}{Hossein~A. Rahmani}, \bibinfo{person}{Daniel Campos}, \bibinfo{person}{Jimmy Lin}, \bibinfo{person}{Ellen~M. Voorhees}, {and} \bibinfo{person}{Ian Soboroff}.} \bibinfo{year}{2024}\natexlab{}.
\newblock \showarticletitle{Overview of the TREC 2023 Deep Learning Track}. In \bibinfo{booktitle}{\emph{Text REtrieval Conference (TREC)}}. NIST, \bibinfo{publisher}{TREC}.
\newblock
\urldef\tempurl%
\url{https://www.microsoft.com/en-us/research/publication/overview-of-the-trec-2023-deep-learning-track/}
\showURL{%
\tempurl}


\bibitem[Dietz et~al\mbox{.}(2022)]%
        {dietz2022automaticwiki}
\bibfield{author}{\bibinfo{person}{Laura Dietz}, \bibinfo{person}{Shubham Chatterjee}, \bibinfo{person}{Connor Lennox}, \bibinfo{person}{Sumanta Kashyapi}, \bibinfo{person}{Pooja Oza}, {and} \bibinfo{person}{Ben Gamari}.} \bibinfo{year}{2022}\natexlab{}.
\newblock \showarticletitle{Wikimarks: Harvesting Relevance Benchmarks from Wikipedia}. In \bibinfo{booktitle}{\emph{Proceedings of the 45th International ACM SIGIR Conference on Research and Development in Information Retrieval}} (Madrid, Spain) \emph{(\bibinfo{series}{SIGIR '22})}. \bibinfo{publisher}{Association for Computing Machinery}, \bibinfo{address}{New York, NY, USA}, \bibinfo{pages}{3003–3012}.
\newblock
\showISBNx{9781450387323}
\urldef\tempurl%
\url{https://doi.org/10.1145/3477495.3531731}
\showDOI{\tempurl}


\bibitem[Dietz and Dalton(2020)]%
        {Dietz2020autotestcollection}
\bibfield{author}{\bibinfo{person}{L. Dietz} {and} \bibinfo{person}{Jeff Dalton}.} \bibinfo{year}{2020}\natexlab{}.
\newblock \showarticletitle{Humans Optional? Automatic Large-Scale Test Collections for Entity, Passage, and Entity-Passage Retrieval.}
\newblock \bibinfo{journal}{\emph{Datenbank-Spektrum}} (\bibinfo{date}{3} \bibinfo{year}{2020}).
\newblock
\showISSN{1618-2162}
\newblock
\shownote{This material is based upon work supported by the National Science Foundation under Grant No. 1846017}.


\bibitem[Eickhoff et~al\mbox{.}(2012)]%
        {eickhoff2012crowd}
\bibfield{author}{\bibinfo{person}{Carsten Eickhoff}, \bibinfo{person}{Christopher~G. Harris}, \bibinfo{person}{Arjen~P. de Vries}, {and} \bibinfo{person}{Padmini Srinivasan}.} \bibinfo{year}{2012}\natexlab{}.
\newblock \showarticletitle{Quality through flow and immersion: gamifying crowdsourced relevance assessments}. In \bibinfo{booktitle}{\emph{Proceedings of the 35th International ACM SIGIR Conference on Research and Development in Information Retrieval}} (Portland, Oregon, USA) \emph{(\bibinfo{series}{SIGIR '12})}. \bibinfo{publisher}{Association for Computing Machinery}, \bibinfo{address}{New York, NY, USA}, \bibinfo{pages}{871–880}.
\newblock
\showISBNx{9781450314725}
\urldef\tempurl%
\url{https://doi.org/10.1145/2348283.2348400}
\showDOI{\tempurl}


\bibitem[Faggioli et~al\mbox{.}(2023)]%
        {Faggioli2023llmreljudge}
\bibfield{author}{\bibinfo{person}{Guglielmo Faggioli}, \bibinfo{person}{Laura Dietz}, \bibinfo{person}{Charles L.~A. Clarke}, \bibinfo{person}{Gianluca Demartini}, \bibinfo{person}{Matthias Hagen}, \bibinfo{person}{Claudia Hauff}, \bibinfo{person}{Noriko Kando}, \bibinfo{person}{Evangelos Kanoulas}, \bibinfo{person}{Martin Potthast}, \bibinfo{person}{Benno Stein}, {and} \bibinfo{person}{Henning Wachsmuth}.} \bibinfo{year}{2023}\natexlab{}.
\newblock \showarticletitle{Perspectives on Large Language Models for Relevance Judgment}. In \bibinfo{booktitle}{\emph{Proceedings of the 2023 ACM SIGIR International Conference on Theory of Information Retrieval}} \emph{(\bibinfo{series}{ICTIR ’23}, Vol.~\bibinfo{volume}{15})}. \bibinfo{publisher}{ACM}, \bibinfo{pages}{39–50}.
\newblock
\urldef\tempurl%
\url{https://doi.org/10.1145/3578337.3605136}
\showDOI{\tempurl}


\bibitem[Ganguly and Yilmaz(2023)]%
        {ganguly2023queryspecificvariabledepthpooling}
\bibfield{author}{\bibinfo{person}{Debasis Ganguly} {and} \bibinfo{person}{Emine Yilmaz}.} \bibinfo{year}{2023}\natexlab{}.
\newblock \bibinfo{title}{Query-specific Variable Depth Pooling via Query Performance Prediction towards Reducing Relevance Assessment Effort}.
\newblock
\newblock
\showeprint[arxiv]{2304.11752}~[cs.IR]
\urldef\tempurl%
\url{https://arxiv.org/abs/2304.11752}
\showURL{%
\tempurl}


\bibitem[Lewis and Gale(1994)]%
        {lewis1994uncertaintysampling}
\bibfield{author}{\bibinfo{person}{David~D. Lewis} {and} \bibinfo{person}{William~A. Gale}.} \bibinfo{year}{1994}\natexlab{}.
\newblock \bibinfo{title}{A Sequential Algorithm for Training Text Classifiers}.
\newblock
\newblock
\showeprint[arxiv]{cmp-lg/9407020}~[cmp-lg]
\urldef\tempurl%
\url{https://arxiv.org/abs/cmp-lg/9407020}
\showURL{%
\tempurl}


\bibitem[Li and Kanoulas(2017)]%
        {li2017activasampling}
\bibfield{author}{\bibinfo{person}{Dan Li} {and} \bibinfo{person}{Evangelos Kanoulas}.} \bibinfo{year}{2017}\natexlab{}.
\newblock \showarticletitle{Active Sampling for Large-scale Information Retrieval Evaluation}. In \bibinfo{booktitle}{\emph{Proceedings of the 2017 ACM on Conference on Information and Knowledge Management}} \emph{(\bibinfo{series}{CIKM ’17})}. \bibinfo{publisher}{ACM}.
\newblock
\urldef\tempurl%
\url{https://doi.org/10.1145/3132847.3133015}
\showDOI{\tempurl}


\bibitem[Losada et~al\mbox{.}(2016)]%
        {losada2016feelinglucky}
\bibfield{author}{\bibinfo{person}{David~E. Losada}, \bibinfo{person}{Javier Parapar}, {and} \bibinfo{person}{\'{A}lvaro Barreiro}.} \bibinfo{year}{2016}\natexlab{}.
\newblock \showarticletitle{Feeling lucky? multi-armed bandits for ordering judgements in pooling-based evaluation}. In \bibinfo{booktitle}{\emph{Proceedings of the 31st Annual ACM Symposium on Applied Computing}} (Pisa, Italy) \emph{(\bibinfo{series}{SAC '16})}. \bibinfo{publisher}{Association for Computing Machinery}, \bibinfo{address}{New York, NY, USA}, \bibinfo{pages}{1027–1034}.
\newblock
\showISBNx{9781450337397}
\urldef\tempurl%
\url{https://doi.org/10.1145/2851613.2851692}
\showDOI{\tempurl}


\bibitem[MacAvaney and Soldaini(2023)]%
        {MacAvaney2023llmreljudge}
\bibfield{author}{\bibinfo{person}{Sean MacAvaney} {and} \bibinfo{person}{Luca Soldaini}.} \bibinfo{year}{2023}\natexlab{}.
\newblock \showarticletitle{One-Shot Labeling for Automatic Relevance Estimation}. In \bibinfo{booktitle}{\emph{Proceedings of the 46th International ACM SIGIR Conference on Research and Development in Information Retrieval}} \emph{(\bibinfo{series}{SIGIR ’23})}. \bibinfo{publisher}{ACM}.
\newblock
\urldef\tempurl%
\url{https://doi.org/10.1145/3539618.3592032}
\showDOI{\tempurl}


\bibitem[Mehrdad et~al\mbox{.}(2024)]%
        {mehrdad2024largelanguagemodelsrelevance}
\bibfield{author}{\bibinfo{person}{Navid Mehrdad}, \bibinfo{person}{Hrushikesh Mohapatra}, \bibinfo{person}{Mossaab Bagdouri}, \bibinfo{person}{Prijith Chandran}, \bibinfo{person}{Alessandro Magnani}, \bibinfo{person}{Xunfan Cai}, \bibinfo{person}{Ajit Puthenputhussery}, \bibinfo{person}{Sachin Yadav}, \bibinfo{person}{Tony Lee}, \bibinfo{person}{ChengXiang Zhai}, {and} \bibinfo{person}{Ciya Liao}.} \bibinfo{year}{2024}\natexlab{}.
\newblock \bibinfo{title}{Large Language Models for Relevance Judgment in Product Search}.
\newblock
\newblock
\showeprint[arxiv]{2406.00247}~[cs.IR]
\urldef\tempurl%
\url{https://arxiv.org/abs/2406.00247}
\showURL{%
\tempurl}


\bibitem[Meng et~al\mbox{.}(2024)]%
        {meng2024queryperformancepredictionusing}
\bibfield{author}{\bibinfo{person}{Chuan Meng}, \bibinfo{person}{Negar Arabzadeh}, \bibinfo{person}{Arian Askari}, \bibinfo{person}{Mohammad Aliannejadi}, {and} \bibinfo{person}{Maarten de Rijke}.} \bibinfo{year}{2024}\natexlab{}.
\newblock \bibinfo{title}{Query Performance Prediction using Relevance Judgments Generated by Large Language Models}.
\newblock
\newblock
\showeprint[arxiv]{2404.01012}~[cs.IR]
\urldef\tempurl%
\url{https://arxiv.org/abs/2404.01012}
\showURL{%
\tempurl}


\bibitem[Nenkova and Passonneau(2004)]%
        {nenkova2004pyramid}
\bibfield{author}{\bibinfo{person}{Ani Nenkova} {and} \bibinfo{person}{Rebecca Passonneau}.} \bibinfo{year}{2004}\natexlab{}.
\newblock \showarticletitle{Evaluating Content Selection in Summarization: The Pyramid Method}. In \bibinfo{booktitle}{\emph{Proceedings of the Human Language Technology Conference of the North {A}merican Chapter of the Association for Computational Linguistics: {HLT}-{NAACL} 2004}}. \bibinfo{publisher}{Association for Computational Linguistics}, \bibinfo{address}{Boston, Massachusetts, USA}, \bibinfo{pages}{145--152}.
\newblock
\urldef\tempurl%
\url{https://aclanthology.org/N04-1019}
\showURL{%
\tempurl}


\bibitem[Oard et~al\mbox{.}(2018)]%
        {doug2018ediscovery}
\bibfield{author}{\bibinfo{person}{Douglas~W. Oard}, \bibinfo{person}{Fabrizio Sebastiani}, {and} \bibinfo{person}{Jyothi~K. Vinjumur}.} \bibinfo{year}{2018}\natexlab{}.
\newblock \showarticletitle{Jointly Minimizing the Expected Costs of Review for Responsiveness and Privilege in E-Discovery}.
\newblock \bibinfo{journal}{\emph{ACM Trans. Inf. Syst.}} \bibinfo{volume}{37}, \bibinfo{number}{1}, Article \bibinfo{articleno}{11} (\bibinfo{date}{Nov.} \bibinfo{year}{2018}), \bibinfo{numpages}{35}~pages.
\newblock
\showISSN{1046-8188}
\urldef\tempurl%
\url{https://doi.org/10.1145/3268928}
\showDOI{\tempurl}


\bibitem[Qin et~al\mbox{.}(2024)]%
        {qin2024llmranking}
\bibfield{author}{\bibinfo{person}{Zhen Qin}, \bibinfo{person}{Rolf Jagerman}, \bibinfo{person}{Kai Hui}, \bibinfo{person}{Honglei Zhuang}, \bibinfo{person}{Junru Wu}, \bibinfo{person}{Le Yan}, \bibinfo{person}{Jiaming Shen}, \bibinfo{person}{Tianqi Liu}, \bibinfo{person}{Jialu Liu}, \bibinfo{person}{Donald Metzler}, \bibinfo{person}{Xuanhui Wang}, {and} \bibinfo{person}{Michael Bendersky}.} \bibinfo{year}{2024}\natexlab{}.
\newblock \bibinfo{title}{Large Language Models are Effective Text Rankers with Pairwise Ranking Prompting}.
\newblock
\newblock
\showeprint[arxiv]{2306.17563}~[cs.IR]
\urldef\tempurl%
\url{https://arxiv.org/abs/2306.17563}
\showURL{%
\tempurl}


\bibitem[Rahman et~al\mbox{.}(2020)]%
        {rahman2020activelearningtestcollection}
\bibfield{author}{\bibinfo{person}{Md~Mustafizur Rahman}, \bibinfo{person}{Mucahid Kutlu}, \bibinfo{person}{Tamer Elsayed}, {and} \bibinfo{person}{Matthew Lease}.} \bibinfo{year}{2020}\natexlab{}.
\newblock \showarticletitle{Efficient Test Collection Construction via Active Learning}. In \bibinfo{booktitle}{\emph{Proceedings of the 2020 ACM SIGIR on International Conference on Theory of Information Retrieval}} (Virtual Event, Norway) \emph{(\bibinfo{series}{ICTIR '20})}. \bibinfo{publisher}{Association for Computing Machinery}, \bibinfo{address}{New York, NY, USA}, \bibinfo{pages}{177–184}.
\newblock
\showISBNx{9781450380676}
\urldef\tempurl%
\url{https://doi.org/10.1145/3409256.3409837}
\showDOI{\tempurl}


\bibitem[Rahmani et~al\mbox{.}(2024)]%
        {rahmani2024llmfortestcollections}
\bibfield{author}{\bibinfo{person}{Hossein~A. Rahmani}, \bibinfo{person}{Nick Craswell}, \bibinfo{person}{Emine Yilmaz}, \bibinfo{person}{Bhaskar Mitra}, {and} \bibinfo{person}{Daniel Campos}.} \bibinfo{year}{2024}\natexlab{}.
\newblock \bibinfo{title}{Synthetic Test Collections for Retrieval Evaluation}.
\newblock
\newblock
\showeprint[arxiv]{2405.07767}~[cs.IR]
\urldef\tempurl%
\url{https://arxiv.org/abs/2405.07767}
\showURL{%
\tempurl}


\bibitem[Roberts et~al\mbox{.}(2021)]%
        {roberts2021treccovid}
\bibfield{author}{\bibinfo{person}{Kirk Roberts}, \bibinfo{person}{Tasmeer Alam}, \bibinfo{person}{Steven Bedrick}, \bibinfo{person}{Dina Demner-Fushman}, \bibinfo{person}{Kyle Lo}, \bibinfo{person}{Ian Soboroff}, \bibinfo{person}{Ellen Voorhees}, \bibinfo{person}{Lucy~Lu Wang}, {and} \bibinfo{person}{William~R Hersh}.} \bibinfo{year}{2021}\natexlab{}.
\newblock \bibinfo{title}{Searching for Scientific Evidence in a Pandemic: An Overview of TREC-COVID}.
\newblock
\newblock
\showeprint[arxiv]{2104.09632}~[cs.IR]
\urldef\tempurl%
\url{https://arxiv.org/abs/2104.09632}
\showURL{%
\tempurl}


\bibitem[Soboroff(2024)]%
        {soboroff2024dontusellms}
\bibfield{author}{\bibinfo{person}{Ian Soboroff}.} \bibinfo{year}{2024}\natexlab{}.
\newblock \bibinfo{title}{Don't Use LLMs to Make Relevance Judgments}.
\newblock
\newblock
\showeprint[arxiv]{2409.15133}~[cs.IR]
\urldef\tempurl%
\url{https://arxiv.org/abs/2409.15133}
\showURL{%
\tempurl}


\bibitem[Stevenson and Bin-Hezam(2023)]%
        {Stevenson2023stopping}
\bibfield{author}{\bibinfo{person}{Mark Stevenson} {and} \bibinfo{person}{Reem Bin-Hezam}.} \bibinfo{year}{2023}\natexlab{}.
\newblock \showarticletitle{Stopping Methods for Technology-assisted Reviews Based on Point Processes}.
\newblock \bibinfo{journal}{\emph{ACM Transactions on Information Systems}} \bibinfo{volume}{42}, \bibinfo{number}{3} (\bibinfo{date}{Dec.} \bibinfo{year}{2023}), \bibinfo{pages}{1–37}.
\newblock
\showISSN{1558-2868}
\urldef\tempurl%
\url{https://doi.org/10.1145/3631990}
\showDOI{\tempurl}


\bibitem[Takehi et~al\mbox{.}(2023)]%
        {Takehi2023nuggetlevel}
\bibfield{author}{\bibinfo{person}{Rikiya Takehi}, \bibinfo{person}{Akihisa Watanabe}, {and} \bibinfo{person}{Tetsuya Sakai}.} \bibinfo{year}{2023}\natexlab{}.
\newblock \showarticletitle{Open-Domain Dialogue Quality Evaluation: Deriving Nugget-level Scores from Turn-level Scores}. In \bibinfo{booktitle}{\emph{Proceedings of the Annual International ACM SIGIR Conference on Research and Development in Information Retrieval in the Asia Pacific Region}} \emph{(\bibinfo{series}{SIGIR-AP ’23})}. \bibinfo{publisher}{ACM}, \bibinfo{pages}{40–45}.
\newblock
\urldef\tempurl%
\url{https://doi.org/10.1145/3624918.3625338}
\showDOI{\tempurl}


\bibitem[Thakur et~al\mbox{.}(2021)]%
        {thakur2021beir}
\bibfield{author}{\bibinfo{person}{Nandan Thakur}, \bibinfo{person}{Nils Reimers}, \bibinfo{person}{Andreas Rücklé}, \bibinfo{person}{Abhishek Srivastava}, {and} \bibinfo{person}{Iryna Gurevych}.} \bibinfo{year}{2021}\natexlab{}.
\newblock \bibinfo{title}{BEIR: A Heterogenous Benchmark for Zero-shot Evaluation of Information Retrieval Models}.
\newblock
\newblock
\showeprint[arxiv]{2104.08663}~[cs.IR]
\urldef\tempurl%
\url{https://arxiv.org/abs/2104.08663}
\showURL{%
\tempurl}


\bibitem[Thomas et~al\mbox{.}(2022)]%
        {thomas2022crowd}
\bibfield{author}{\bibinfo{person}{Paul Thomas}, \bibinfo{person}{Gabriella Kazai}, \bibinfo{person}{Ryen White}, {and} \bibinfo{person}{Nick Craswell}.} \bibinfo{year}{2022}\natexlab{}.
\newblock \showarticletitle{The Crowd is Made of People: Observations from Large-Scale Crowd Labelling}. In \bibinfo{booktitle}{\emph{Proceedings of the 2022 Conference on Human Information Interaction and Retrieval}} (Regensburg, Germany) \emph{(\bibinfo{series}{CHIIR '22})}. \bibinfo{publisher}{Association for Computing Machinery}, \bibinfo{address}{New York, NY, USA}, \bibinfo{pages}{25–35}.
\newblock
\showISBNx{9781450391863}
\urldef\tempurl%
\url{https://doi.org/10.1145/3498366.3505815}
\showDOI{\tempurl}


\bibitem[Thomas et~al\mbox{.}(2024)]%
        {thomas2024llmreljudge}
\bibfield{author}{\bibinfo{person}{Paul Thomas}, \bibinfo{person}{Seth Spielman}, \bibinfo{person}{Nick Craswell}, {and} \bibinfo{person}{Bhaskar Mitra}.} \bibinfo{year}{2024}\natexlab{}.
\newblock \bibinfo{title}{Large language models can accurately predict searcher preferences}.
\newblock
\newblock
\showeprint[arxiv]{2309.10621}~[cs.IR]
\urldef\tempurl%
\url{https://arxiv.org/abs/2309.10621}
\showURL{%
\tempurl}


\bibitem[Upadhyay et~al\mbox{.}(2024a)]%
        {upadhyay2024llmspatchmissingrelevance}
\bibfield{author}{\bibinfo{person}{Shivani Upadhyay}, \bibinfo{person}{Ehsan Kamalloo}, {and} \bibinfo{person}{Jimmy Lin}.} \bibinfo{year}{2024}\natexlab{a}.
\newblock \bibinfo{title}{LLMs Can Patch Up Missing Relevance Judgments in Evaluation}.
\newblock
\newblock
\showeprint[arxiv]{2405.04727}~[cs.IR]
\urldef\tempurl%
\url{https://arxiv.org/abs/2405.04727}
\showURL{%
\tempurl}


\bibitem[Upadhyay et~al\mbox{.}(2024b)]%
        {upadhyay2024largescalestudyrelevanceassessments}
\bibfield{author}{\bibinfo{person}{Shivani Upadhyay}, \bibinfo{person}{Ronak Pradeep}, \bibinfo{person}{Nandan Thakur}, \bibinfo{person}{Daniel Campos}, \bibinfo{person}{Nick Craswell}, \bibinfo{person}{Ian Soboroff}, \bibinfo{person}{Hoa~Trang Dang}, {and} \bibinfo{person}{Jimmy Lin}.} \bibinfo{year}{2024}\natexlab{b}.
\newblock \bibinfo{title}{A Large-Scale Study of Relevance Assessments with Large Language Models: An Initial Look}.
\newblock
\newblock
\showeprint[arxiv]{2411.08275}~[cs.IR]
\urldef\tempurl%
\url{https://arxiv.org/abs/2411.08275}
\showURL{%
\tempurl}


\bibitem[Voorhees(1998)]%
        {voorhees1998overlap}
\bibfield{author}{\bibinfo{person}{Ellen~M. Voorhees}.} \bibinfo{year}{1998}\natexlab{}.
\newblock \showarticletitle{Variations in relevance judgments and the measurement of retrieval effectiveness}. In \bibinfo{booktitle}{\emph{Proceedings of the 21st Annual International ACM SIGIR Conference on Research and Development in Information Retrieval}} (Melbourne, Australia) \emph{(\bibinfo{series}{SIGIR '98})}. \bibinfo{publisher}{Association for Computing Machinery}, \bibinfo{address}{New York, NY, USA}, \bibinfo{pages}{315–323}.
\newblock
\showISBNx{1581130155}
\urldef\tempurl%
\url{https://doi.org/10.1145/290941.291017}
\showDOI{\tempurl}


\bibitem[Voorhees(2005)]%
        {voorhees2005robust}
\bibfield{author}{\bibinfo{person}{Ellen~M. Voorhees}.} \bibinfo{year}{2005}\natexlab{}.
\newblock \showarticletitle{The TREC robust retrieval track}.
\newblock \bibinfo{journal}{\emph{SIGIR Forum}} \bibinfo{volume}{39}, \bibinfo{number}{1} (\bibinfo{date}{jun} \bibinfo{year}{2005}), \bibinfo{pages}{11–20}.
\newblock
\showISSN{0163-5840}
\urldef\tempurl%
\url{https://doi.org/10.1145/1067268.1067272}
\showDOI{\tempurl}


\bibitem[Voorhees(2018)]%
        {voorhees2018bandit}
\bibfield{author}{\bibinfo{person}{Ellen~M. Voorhees}.} \bibinfo{year}{2018}\natexlab{}.
\newblock \showarticletitle{On Building Fair and Reusable Test Collections using Bandit Techniques}. In \bibinfo{booktitle}{\emph{Proceedings of the 27th ACM International Conference on Information and Knowledge Management}} (Torino, Italy) \emph{(\bibinfo{series}{CIKM '18})}. \bibinfo{publisher}{Association for Computing Machinery}, \bibinfo{address}{New York, NY, USA}, \bibinfo{pages}{407–416}.
\newblock
\showISBNx{9781450360142}
\urldef\tempurl%
\url{https://doi.org/10.1145/3269206.3271766}
\showDOI{\tempurl}


\bibitem[Voorhees et~al\mbox{.}(2022)]%
        {voorhees2022relevants}
\bibfield{author}{\bibinfo{person}{Ellen~M. Voorhees}, \bibinfo{person}{Nick Craswell}, {and} \bibinfo{person}{Jimmy Lin}.} \bibinfo{year}{2022}\natexlab{}.
\newblock \showarticletitle{Too Many Relevants: Whither Cranfield Test Collections?}. In \bibinfo{booktitle}{\emph{Proceedings of the 45th International ACM SIGIR Conference on Research and Development in Information Retrieval}} (Madrid, Spain) \emph{(\bibinfo{series}{SIGIR '22})}. \bibinfo{publisher}{Association for Computing Machinery}, \bibinfo{address}{New York, NY, USA}, \bibinfo{pages}{2970–2980}.
\newblock
\showISBNx{9781450387323}
\urldef\tempurl%
\url{https://doi.org/10.1145/3477495.3531728}
\showDOI{\tempurl}


\bibitem[Voorhees and Harman(2000)]%
        {voorhees2000overview}
\bibfield{author}{\bibinfo{person}{Ellen~M. Voorhees} {and} \bibinfo{person}{Donna Harman}.} \bibinfo{year}{2000}\natexlab{}.
\newblock \showarticletitle{Overview of the Eighth Text {REtrieval} Conference ({TREC-8})}. In \bibinfo{booktitle}{\emph{Proceedings of the Eighth Text {REtrieval} Conference ({TREC-8})}} \emph{(\bibinfo{series}{NIST Special Publication}, \bibinfo{number}{500-246})}, \bibfield{editor}{\bibinfo{person}{E.M. Voorhees} {and} \bibinfo{person}{D.K. Harman}} (Eds.). \bibinfo{pages}{1--24}.
\newblock


\end{thebibliography}
\end{document}